\documentclass[12pt]{article}
\usepackage{epsfig,amssymb}

\newcommand{\bq}{\begin{equation}}
\newcommand{\eq}{\end{equation}}
\newcommand{\bqa}{\begin{eqnarray}}
\newcommand{\eqa}{\end{eqnarray}}
\newcommand{\ben}{\begin{enumerate}}
\newcommand{\een}{\end{enumerate}}
\newcommand{\bc}{\begin{center}}
\newcommand{\ec}{\end{center}}
\newcommand{\bqb}{\begin{eqnarray*}}
\newcommand{\eqb}{\end{eqnarray*}}

\def\ie{{\it i.e. }}
\def\eg{{\it e.g. }}

\def\etal{{\it et.al. }}

\def\L{ {\cal L }}
\def\tF{ {\tilde F }}
\def\B{\tilde {\cal B}}
\def\Del{\tilde { \Delta}}
\def\sw{s_W}
\def\cw{c_W}
\def\swd{s^2_W}
\def\cwd{c^2_W}

\def\mwd{m_W^2}
\def\mw{m_W}
\def\mz{m_Z}
\def\mzd{m_Z^2}
\def\mt{m_t}

\def\mchil{m_{\tilde \chi_1}}
\def\mchih{m_{\tilde \chi_2}}

\def\Sn#1{\mathrm{Sign} #1 }

\def\tchi{\tilde \chi}
\def\stop{\tilde t}
\def\cfR{\cos\phi_R}
\def\cfL{\cos\phi_L}
\def\sfR{\sin\phi_R}
\def\sfL{\sin\phi_L}
\def\calpha{\cos \alpha}
\def\cbeta{\cos\beta}
\def\salpha{\sin \alpha}
\def\sbeta{\sin \beta}

\def\pr#1#2#3{ Phys. Rev. ${\bf{#1}}$, #2 (#3)}

\def\pl#1#2#3{ Phys. Lett. ${\bf{#1}}$, #2 (#3)}
\def\prep#1#2#3{ Phys. Rep. ${\bf{#1}}$, #2 (#3)}

\def\np#1#2#3{ Nucl. Phys. ${\bf{#1}}$, #2 (#3)}
\def\zp#1#2#3{ Z. f. Phys. ${\bf{#1}}$, #2 (#3)}
\def\epj#1#2#3{ Eur. Phys. J. ${\bf{#1}}$, #2 (#3)}

\def\cpc#1#2#3{Comput. Phys. Commun. ${\bf{#1}}$, #2 (#3)}

\begin{document}
\pagenumbering{arabic}
\thispagestyle{empty}
\def\thefootnote{\fnsymbol{footnote}}
\setcounter{footnote}{1}

\begin{flushright}
PM/01-12 \\
THES-TP 2001/05 \\
March 2001\\
 \end{flushright}
\vspace{2cm}
\begin{center}
{\Large\bf The processes
$\gamma\gamma\to Z H $ in SM
and  MSSM\footnote{Partially
supported by EU contract  HPRN-CT-2000-00149.}.}
 \vspace{1.5cm}  \\
{\large G.J. Gounaris$^a$, P.I. Porfyriadis$^a$ and
F.M. Renard$^b$}\\
\vspace{0.7cm}
$^a$Department of Theoretical Physics, Aristotle
University of Thessaloniki,\\
Gr-54006, Thessaloniki, Greece.\\
\vspace{0.2cm}
$^b$Physique
Math\'{e}matique et Th\'{e}orique,
UMR 5825\\
Universit\'{e} Montpellier II,
 F-34095 Montpellier Cedex 5.\\
\vspace{0.2cm}

 \vspace*{1cm}

{\bf Abstract}
\end{center}

The process $\gamma\gamma\to ZH$ first arises at the 1-loop level,
and as such it provides
remarkable  tests of  the structure of the electroweak
 Higgs sector.
These tests are complementary to
those in the gauge sector involving  $\gamma\gamma\to
\gamma\gamma,~\gamma Z,~ZZ$. We show that in the Standard Model (SM)
 where  $H=H_{SM}$, as well as in the  supersymmetric  case where
$H=h^0,~H^0$ or $A^0$,
 observables exist (like \eg the energy
dependence, angular distribution, photon polarization dependence or
final $Z$ polarization) which present rather spectacular
properties. Such   properties involve
strong threshold effects with steps, bumps or peaks,  reflecting the
type of Higgs and heavy quarks  and
  chargino masses and couplings predicted
by  the SM and supersymmetric models.

\def\thefootnote{\arabic{footnote}}
\setcounter{footnote}{0}
\clearpage

\section{Introduction}

\hspace{0.6cm}Photon-photon collisions have been recognized as being
a remarkable place for testing the structure of the electroweak
interactions at high energy, both in the gauge and in the Higgs sector
\cite{ggrev}. These collisions should be experimentally feasible
with the high intensity achievable through the laser backscattering
procedure at a linear $e^+e^-$ collider \cite{laser}.
Many such studies \cite{ggstudies}
have been done in connection with the $e^+e^-$ collider
projects LC \cite{LC} and CLIC \cite{CLIC}.

The significance of the photon-photon processes stems from the fact
that they provide new tests of the fundamental interactions, which are
often complementary to those achievable in direct $e^+e^-$ collisions.
These consist either in precise measurements sensitive to
high order effects among standard and new particles,
or in independent ways of producing new particles.

Of particular importance is the experimental study of
the Higgs sector of
the electroweak interactions, for which the Standard
Model (SM) and the various
extended models, like \eg the Minimal Supersymmetric
Standard Model (MSSM), give specific examples.
In this respect, the basic photon-photon process is
$\gamma\gamma\to H$, where $H$ is a standard or a non standard
neutral Higgs boson. This process arises at 1-loop and provides
interesting tests of the Higgs boson couplings to the particles
running inside the triangle loop; which could be the
standard gauge bosons, leptons
and quarks, as well as any new charged particles
 that might exist. New Higgs interactions could also be searched
this way \cite{Hgg}.

However the information obtained from $\gamma\gamma\to H$
is restricted by the kinetic  constraint
$s=m^2_H$. To go beyond this, it is natural to look
at the associate production $\gamma\gamma\to ZH$ in which
several observables sensitive to the dynamical contents,
 may be accessible. In SM or SUSY models, such
processes   first arise at the one loop level, contrary  to
the complementary process $e^+e^-\to ZH$ which is dominated
by the tree level contribution involving the $ZZH$ coupling.
So $\gamma\gamma\to ZH$, which has many similarities with
the previously studied processes $\gamma\gamma\to
\gamma\gamma,~\gamma Z,~ZZ$
\cite{Jikia, gggg,gggZ,ggZZ,ggZZ-2nd} , should be sensitive to the
quantum effects of the scalar sector and to
the Higgs boson interactions with the particles running
inside the loops.

In this paper we consider therefore the process $\gamma\gamma\to ZH$
where $H$ is either the
standard Higgs boson $H_{SM}$, or a supersymmetric $h^0,H^0$ or
$A^0$ state.

The dynamical contents at one loop
is rather simple, but physically important.
The generic form of the Feynman
diagrams is depicted in Figs.1,2. It  consists of
triangle diagrams related to either an intermediate Higgs
boson in the s-channel, or to a $Z$ (plus Goldstone $G^0$)
exchange; and of box diagrams. These we classify as follows:
\par

a) The diagrams with an intermediate Higgs boson in the s-channel
 only exist in the SUSY cases $\gamma\gamma \to A^0 \to Zh^0,~ ZH^0$
 and $\gamma\gamma \to H^0,h^0\to ZA^0$;
see Fig.\ref{ggZh0-diag}a  for an $A^0$ exchange and
 Figs.\ref{ggZA0-diag}a respectively.
The related triangular loops describing
$\gamma\gamma \to H^0,h^0,A^0$ have been studied before and
involve standard and supersymmetric bosonic and fermionic
loops. These contributions are especially important in the
$\gamma\gamma \to A^0\to Zh^0$ case for energies close to the
$A^0$ pole.\par

b) The diagrams with a  $(Z,~G^0)$-exchange  involve the anomalous
$Z\gamma\gamma$  and $G^0\gamma\gamma$ fermionic triangles,
and the final $(Z,G^0)ZH_{SM}$, $(Z,G^0)Zh^0$ and
$(Z,G^0)ZH^0$ couplings; see\footnote{ Notice that there is no
such contribution for $ZA^0$ production.}
Fig.\ref{ggZh0-diag}a,b. This contribution vanishes when the
$Z$ is on shell, forcing the  whole term to behave
 like a contact interaction with vanishing total angular momentum
in the s-channel.\par

c) The box diagrams always involve fermionic loops; see Fig.1c-g,
Fig.2b-f. No bosonic loop is allowed because of the charge
conjugation properties
of the boson couplings. In SM, the fermionic boxes only
involve the standard lepton
and quark contributions. The top quark contribution
is  predominant in this case, because of the two fermion mass
factors imposed respectively on the amplitude
by the Higgs couplings and  the chirality violating nature of the
process. In SUSY, for sufficient
large $\tan\beta$, the importance of all  quarks and leptons of the
third family may be comparable; and we have in addition
chargino boxes, involving either a single chargino
running along the loop, or both  charginos; (the later
 we call mixed chargino contribution).

The purpose of our study is to see  how the various parts
of the above contents
reflect on the properties of the process $\gamma\gamma\to ZH$,
and how this may  be useful in testing the SM and  MSSM models.

The contents of the paper is the following.
In Section 2, we collect the elements of the SM and MSSM Lagrangian
needed to compute the amplitudes in the four cases
$H=H_{SM}, h^0,H^0,A^0$. The various  couplings are
collected in Appendix A. The helicity amplitudes generated by
the various  diagrams are
explicitly given in analytic form in the Appendices
B and C.
In Section 3 we discuss the properties of the various observables
of the process $\gamma\gamma \to
ZH$. We consider the unpolarized and polarized
$\gamma\gamma$ cross sections, the $ZH$ angular distributions
and the final $Z$ polarization. Several illustrations are
given for  SM and  MSSM. A summary of the results is made
in Section 4.

\vspace{1cm}
\section{Dynamical characteristics of the process
$\gamma\gamma\to ZH$}

The generic set of the contributing
 diagrams is depicted in Fig.1a-g for the cases of
$\gamma\gamma\to Zh^0$ and $\gamma\gamma\to ZH^0$; and in Fig.2a-f for
the case of $\gamma\gamma\to ZA^0$. The SM case
$\gamma\gamma\to ZH_{SM}$ is obtained from
Fig.1 by retaining only diagrams (1c) and (1b), together with the
Goldstone involving part of  (1a).

Boson loop contributions
can only appear in the triangle diagram in Fig.\ref{ggZA0-diag}a,
and involve $W^{\pm}$ (plus goldstone and  ghost) and
 charged Higgs,  charged sleptons and squark
lines. Their contributions have already been computed previously
\cite{ggZZ-2nd} and of course affects only
$\gamma \gamma \to Z A^0$.

In all other diagrams, only internal fermion lines occur.
These are the triangle diagrams Fig.1a,b and 2a, and the box diagrams
1c, 2b, involving  internal standard charged fermion lines
(leptons and quarks), as well as single chargino lines.
Our conventions for the  gauge boson couplings
and  the Yukawa couplings of the Higgs bosons
($H^{SM},h^0,H^0,A^0$) to
leptons and quarks are  given in Appendix A. Note that
the Yukawa couplings depend on the SUSY parameters $\alpha$
and $\beta$ of the Higgs sector, for which our conventions are as
in \cite{ggZZ-2nd}.

The contribution of the third family of quarks and leptons,
(essentially only the top quark in SM or low $\tan\beta$ SUSY models),
is strongly dominating the one from the lighter quarks and
leptons. The reason is due to the presence of a factor $m_f$
in the Yukawa couplings on the one hand; and due to  the
chirality flip along the fermionic lines of  the loop,
which introduces an additional $m_f$ factor.

The $Z,G^0$ exchange contribution
corresponding to the diagrams (1b,a)
has no $Z$-pole factor, and
behaves like a contact interaction with the quantum numbers
of a scalar exchange in the s-channel. It turns out that it is
quite important in all SM or MSSM cases.

As already stated, the diagrams in Figs.1a-c, 2a.b also
describe the contributions from a single chargino
$\tilde{\chi}^{\pm}$ running along
the loop. Since the Yukawa-type couplings of the charginos
involve no masses though, there is one power  of fermion
masses less, compared to the ($t,~b,~\tau$) case; see
(\ref{H-chi1-coupling}, \ref{H-chi2-coupling}).

In addition to them though, we have the box diagrams Fig.1d-g,
Fig.2c-f ($j\neq i$) involving
mixed chargino lines, due to the possibility
of mixed $Z\tilde{\chi_1}\tilde{\chi_2}$
and $H\tilde{\chi_1}\tilde{\chi_2}$
couplings. The various unmixed and mixed couplings are
defined in (\ref{Z-chi1-coupling}-\ref{Z-chi12-coupling},
\ref{H-chi1-coupling}-\ref{H-chi12-coupling}).
They involve the full set of parameters of the SUSY chargino
sector \cite{ggZZ-2nd}.

We have computed the helicity amplitudes
$F_{\lambda_1, \lambda_2, \lambda_Z}$ of the
$\gamma\gamma\to ZH$ process ($H=H^{SM},h^0,H^0,A^0$)
generated by all these diagrams.
They are explicitly given in Appendix B for the
$H_{SM},~h^0, ~H^0$ production cases, and in C for the $A^0$
case\footnote{For their definitions see
(\ref{ggZh0-process}, \ref{ggZA0-process})}.
The expressions are
in terms of the Passarino-Veltman functions
($C_0,~ D_0$) functions.
As explained in the Appendices B and C, owing to the
CP-invariance and Bose symmetry, there are only
four "basic" amplitudes
\bq
F_{+++}~~~,~~~ F_{+--}~~~,~~~ F_{++0}~~~,~~~ F_{+-0} ~~~ ,
\label{basic-amplitudes}
\eq
compare (\ref{basic-h0-amplitudes}, \ref{basic-A0-amplitudes}),
from which all the other ones can be obtained.
See also (\ref{ampl-con1}, \ref{CP-h0-con}, \ref{CP-A0-con}),
 as well as (\ref{ampl-con2}, \ref{ampl-con4}).

Before computing the various    observables,  we should  point out
 certain  important properties
of the 1-loop contributions to the $\gamma\gamma\to ZH$
helicity amplitudes.

Because of the scalar or pseudoscalar nature of the intermediate
state, the triangle diagrams connected either to an intermediate
Higgs boson or to an intermediate (virtual) $Z,G^0$-exchange,
contribute  only to the $F_{\pm\pm0}$ amplitude;
compare the diagrams in Figs.\ref{ggZh0-diag}a,b,
and \ref{ggZA0-diag}a.

The (fermionic) box diagrams also favor   the dominance of the
$F_{\pm\pm0}$ amplitude. This is due to the
chirality violating Higgs-fermion coupling on the one hand,
 and the Bose statistics for the two initial photons on the other.
The chirality argument goes as follows.
When the intermediate fermion-antifermion
state is physical, chirality violation means
$\lambda_f=\lambda_{\bar f}$
for the fermion and antifermion helicities, which then favors
$\lambda_Z=0$;  i.e. dominance of longitudinal $Z$ production.
In addition to it, Eqs. (\ref{ampl-con3}, \ref{ampl-con4}),
imposed by Bose symmetry, lead to the  expectation
that $|F_{\pm\pm 0}| \gg |F_{\pm\mp 0}|$ at large angles. \par

We expect therefore that the whole contribution to
the process $\gamma\gamma\to ZH$
should be dominated by the $F_{\pm\pm 0}$ amplitude.
In a photon-photon collider this
dominance of $Z_LH$ production could be tested
by looking
at the decay distribution $Z\to f\bar f$, especially if one
could study the
charged lepton pairs. Moreover, the dominance of the
$\Delta\lambda=0$ amplitudes should lead to
a very simple form for the polarized photon-photon cross section

We next turn to the numerical results which indeed confirm the
above expectations.

\vspace{1cm}
\section{Results for the observables
of the process $\gamma \gamma \to ZH$}

In a $\gamma\gamma$ Collider generated through Laser
backscattering and employing various polarizations of laser photons
and the $e^\pm$ beams, we can a priori measure various types of
"cross sections" through  \cite{gggg,gggZ,ggZZ}
\bqa
{d\sigma\over d\tau d\cos\vartheta}&=&{d \bar L_{\gamma\gamma}\over
d\tau} \Bigg \{
{d\bar{\sigma}_0\over d\cos\vartheta}
+\langle \xi_2 \xi_2^\prime
\rangle{d\bar{\sigma}_{22}\over d\cos\vartheta}
+\langle\xi_3 \xi_3^\prime\rangle
{d\bar{\sigma}^\prime_{33}\over d\cos\vartheta}
\cos2(\phi- \phi^\prime) + ... \Bigg \}~ ,
\label{sigLCgg}
\eqa
where the dots stand for the various
 "cross section" $\bar \sigma_j$ which do not involve the
large $F_{\pm\pm0}(\gamma \gamma \to H Z)$ amplitudes.
In (\ref{sigLCgg}), $\tau=s /s_{ee}$ as usual, where
$s\equiv s_{\gamma \gamma}$ is defined in (\ref{kin1}), while
$ d \bar L_{\gamma\gamma}/d\tau$ describes the photon-photon
luminosity per unit $e^-e^+$ flux
\cite{ggrev, laser, ggstudies}. The Stokes parameters
$(\xi_2, \xi_2^\prime)$, $(\xi_3,~ \xi_3^\prime)$
and $(\phi,~ \phi^\prime)$ describe respectively
 the average helicities, transverse
polarizations and azimuthal angles of the two
 backscattered photons \cite{gggg,gggZ,ggZZ}). In
(\ref{sigLCgg}) there appear the "cross section" quantities
\bqa
{d\bar \sigma_0(\gamma \gamma \to ZH) \over d\cos\vartheta}& \equiv &
{\kappa \over 64 \pi s^2}
\sum_{\lambda_Z} \left [|F_{++\lambda_Z}|^2
+|F_{+-\lambda_Z}|^2 \right ] ~ ,  \label{sig0} \\
{d\bar{\sigma}_{22}(\gamma \gamma \to HZ)\over d\cos\vartheta} &\equiv&
{\kappa \over 64 \pi s^2} \sum_{\lambda_Z}
\left [|F_{++\lambda_Z}|^2
-|F_{+-\lambda_Z}|^2 \right ]  \ , \label{sig22} \\
{d\bar{\sigma}^\prime_{33}(\gamma \gamma \to HZ)\over d\cos\vartheta}
&\equiv&
{\kappa \over 64 \pi s^2 } \sum_{\lambda_Z}
Re[F_{++\lambda_Z}F^*_{--\lambda_Z}] \  ,
\label{sig33prime}
\eqa
where  $\kappa$ is  defined in (\ref{kin1}) and it is related to the
common $Z$ and   $H$ momenta in their c.m. frame   through
(\ref{kin2}); while $\vartheta$ is the scattering angle in the
same frame. Notice that $\bar \sigma_0$ is the unpolarized
$\gamma \gamma \to Z H$ cross section.
If  only the  $F_{\pm\pm0}$ amplitude were
retained in (\ref{sig0}-\ref{sig33prime}), we would had gotten
\bqa
{d\bar \sigma_0\over d\cos\vartheta }\simeq
{d\bar{\sigma}_{22}\over d\cos\vartheta}\simeq
\eta{d\bar{\sigma}^\prime_{33}\over d\cos\vartheta}
\simeq
\left ({\kappa \over64\pi s^2}\right )
|F_{++0}|^2 ~~ ,   \label{sig-aprox}
\eqa
with $\eta=-1$ for $H_{SM},h^0,H^0$ and $\eta=+1$ for $A^0$
\cite{gggg}; compare (\ref{CP-h0-con}, \ref{CP-A0-con}).
These simple expressions imply very clean tests of the absence of
unexpected contribution (beyond SM or MSSM), to be performed
using polarized laser and $e^{\pm}$ beams.
We now discuss separately the 4 cases of neutral Higgs
boson production.

\vspace{0.5cm}
\noindent
{\bf The SM case.}\\
In Fig.\ref{ggZH-SM-fig}a, we present the SM results
for the $\bar \sigma_0, ~\bar \sigma_{22}$   "cross sections"
integrated in the region $\pi/3 \leq \vartheta \leq 2\pi/3$,
after summing over all $Z$-polarizations.
We use\footnote{Here, as well as in \cite{ggZZ-2nd}, we  use
$\alpha=1/137$. This is to be contrasted to the  results
in \cite{gggg, gggZ, ggZZ} where $\alpha=1/128$ was used causing an
increase of the overall magnitude of the various
cross sections due to their $\alpha^2$ factor.}  $m_H=130GeV$.
In Fig.\ref{ggZH-SM-fig}b the corresponding differential cross
sections are given for the cases of $Z$ production, either  with
all possible Z-polarizations summed, or with just
$\lambda_Z=0$ retained.

As can be seen in Fig.\ref{ggZH-SM-fig}a,b,
the differential and total "cross sections" for
$\bar\sigma_0$ and $\bar \sigma_{22}$ are almost identical, and
also equal to the corresponding cross sections for longitudinal
$Z$-production. In fact we find that (\ref{sig-aprox}) is very
accurately satisfied for all scattering
angles, which just confirms that $F_{\pm\pm 0}$
very strongly dominates all other amplitudes   in the SM case.

In Fig.\ref{ggZH-SM-fig}a, a spectacular peak appears
at the $t\bar t$ threshold, which  comes from the top quark
contribution to the box-diagrams, as well as to the triangle ones
inducing the  anomalous $Z,G^0$ contributions.
It turns out that these  contributions have similar
sizes and  interfere destructively at high energy, thus enforcing
the fast decrease of the cross section.
The angular distribution (see Fig.\ref{ggZH-SM-fig}b,
paying attention to the scale in the y-axis) is, as expected from
the relevant diagrams, rather flat. This may allow a clean
detection of the $ZH$ final state at large angles.

\vspace{0.5cm}
\noindent
{\bf The MSSM cases}\\
We next turn to the  supersymmetric
cases of  $h^0, ~ H^0, ~A^0$ production, exploring various
sets of SUSY parameters. Two extreme typical
sets with $\tan\beta=5$ (set A)
and $\tan\beta=50$ (set B) are illustrated in
Figs.\ref{ggZh0-SUSY-fig}-\ref{ggZA0-SUSY-fig}. The corresponding
parameters,  were calculated employing the unification condition
\bq
M_1=\frac{5}{3}\tan^2\theta_W M_2 ~~,
\eq
and using the HDECAY code  \cite{HDECAY}. The results for
 the physical masses and
widths of the various Higgs bosons, the $(\stop_1,~\stop_2)$-squarks and
the charginos,  are presented\footnote{We have checked that the parameters
in Sets A and B satisfy the requirements for the absence of
charge or colour braking \cite{LeMouel}} in Table 1. In the calculations
of the loops in all SUSY examples below, we just retain the quarks
and leptons of the third family, the charginos, the gauge-bosons
(together with their associated goldstone bosons and ghosts), and
the charged Higgs and $\stop_1,~\stop_2$ bosons.

\begin{table}[htb]
\begin{center}
{ Table 1: SUSY Examples. \\
(Particle masses and widths at the electroweak scale.) }\\
\vspace*{0.3cm}
\begin{tabular}{||c|c|c||}
\hline \hline
\multicolumn{3}{||c||}{$M_2=200 \rm GeV$~ ,~ $\mu=300 \rm GeV$ ~,
 ~ $M_{\tilde f} \simeq 1000 \rm GeV $ }
\\ \hline
 & Set A & Set B  \\ \hline
\multicolumn{1}{||c|}{$\tan \beta $ } &
\multicolumn{1}{|c|}{ 5 }&
\multicolumn{1}{|c||}{ 50 }
\\
\multicolumn{1}{||c|}{$A_t=A_b=A_\tau$ (GeV)}&
\multicolumn{1}{|c|}{2550 }&
\multicolumn{1}{|c||}{ 2500}
\\
\multicolumn{1}{||c|}{$m_{\stop_1}$ (GeV)}&
\multicolumn{1}{|c|}{ 781 }&
\multicolumn{1}{|c||}{780 }
\\
\multicolumn{1}{||c|}{$m_{\stop_2}$ (GeV)}&
\multicolumn{1}{|c|}{ 1201 }&
\multicolumn{1}{|c||}{1201}
\\
\multicolumn{1}{||c|}{$m_{\tchi_1} $(GeV) }&
\multicolumn{1}{|c|}{ 170 }&
\multicolumn{1}{|c||}{180}
\\
\multicolumn{1}{||c|}{$m_{\tchi_2}$ (GeV)}&
\multicolumn{1}{|c|}{ 337 }&
\multicolumn{1}{|c||}{333}
\\
$m_{h^0}$(GeV)  & 119  & 126    \\
$\Gamma_{h^0}$(GeV)  & 0.0089   & 0.049      \\
$m_{H^0 }$ (GeV) & 205  & 150   \\
$\Gamma_{H^0}$(GeV)  & 0.135  & 8.02 \\
 $m_{A^0}$ (GeV) & 200  & 150     \\
$\Gamma_{A^0}$(GeV)  & 0.114  & 8.08 \\
$m_{H^\pm}$ (GeV) & 215  & 168    \\
 \hline \hline
\end{tabular}
\end{center}
\end{table}

As one sees from the differential cross sections in  these
Figures, the dominance of $Z_L$ production
is true in all cases at the level of more than $98\%$.
Also the equality of $\bar{\sigma}_0$ with $\bar{\sigma}_{22}$ (and also
with $\bar{\sigma}_{33}^\prime$ not shown in this figure)
is effective for $h^0$ and $A^0$ at more than $98\%$,
and for $H^0$ at more than $95\%$. We have checked that these
results remain true as we go down in energy approaching the
production threshold.

We now add specific comments for each of the supersymmetric
Higgs bosons.

\vspace{0.5cm}
\noindent
{\bf $h^0$ production}.\\
As expected from the similarity of the basic $h^0$ and $H_{SM}$
couplings, this case is
very close to the SM one. This is confirmed by the comparison
of Figs.\ref{ggZh0-SUSY-fig} and \ref{ggZH-SM-fig}.
As is shown in Fig.\ref{ggZh0-SUSY-fig}a,c,  there is
a strong dominance of the top quark box contribution
and a large contribution from the anomalous $Z,G^0$-exchange
diagrams, like in the SM case. For the case of Set B in particular
(Fig.\ref{ggZh0-SUSY-fig}c), the large $\tan\beta$ value
implies also  appreciable  $b$-quark and $\tau$-lepton contributions,
which somewhat  enhance the magnitude of the cross sections, compared to
those of Set A. The chargino box contributes
at most $10\%$  of the cross section, and produces only
small modifications around the two chargino thresholds.
The angular distribution is also similar to the SM one.

\vspace{0.5cm}
\noindent
{\bf $H^0$ production}.\\
The results for the parameter Sets A and B of Table 1 are
shown in Figs.\ref{ggZH0-SUSY-fig}a-d. In this case there is
no important   top  quark contribution to the
box and  to  the  anomalous $Z,G^0$ diagrams, because the
$H^0t \bar t$ coupling is weaker than the $h^0 t \bar t$ one,
and decreasing as $\tan\beta$ increases \cite{Milada}.
This reduces considerably the $H^0$ production cross sections, as
compared to the  $h^0$-ones. But at the same time, it allows for the
appearance of very strong threshold effects due to the chargino boxes.

The shape and the size of these effects depend  directly on
the choice of the MSSM parameters controlling the size of the
$H \tilde{\chi}_i \tilde{\chi}_j$ couplings.
The result is a rather complex
addition of unmixed and mixed chargino contributions.
Sets A and B illustrate how one can get
steps or peaks depending on the phase of the box amplitude
(the relative size of the real and imaginary parts around the
threshold) interfering with the real and imaginary parts of the
$t\bar t$ box. Steps are essentially due to the imaginary parts,
while  peaks are due to the real parts.
So one has here a very nice way of testing
the choice of MSSM parameters.
The angular distribution is also rather flat but, depending on the set
of SUSY parameters, one can see small violations of the
$\bar{\sigma}_0=\bar{\sigma}_{22}=\bar{\sigma}'_{33}$ rule.
So in this process the chargino contribution is very
important and lead to several kinds of typical effects.

\vspace{0.5cm}
\noindent
{\bf $A^0$ production}.\\
This $A^0$-production case, illustrated in Fig.\ref{ggZA0-SUSY-fig}a-d,
is somewhat different from  the $h^0$ and the
$H^0$ ones, because of the absence of anomalous $Z,G^0$
contribution (there are no $ZZA^0$ and $G^0ZA^0$ couplings),
and because the size of the $A^0t \bar t$ and
$A^0\chi^+\chi^-$ couplings is different from the $h^0$ or $H^0$ ones.
The contribution  of the $t$-quark box is less pronounced than
for $h^0$, but larger than for $H^0$. Correspondingly the
chargino threshold effects have different shapes, i.e. steps or
large bumps instead of narrow peaks. The sensitivity to the choice of
MSSM parameters is still very large; compare the set A and B results
 illustrated in Fig.\ref{ggZA0-SUSY-fig}a-d.
The angular distributions of the various cross sections are always
rather flat, but different curvatures  appear,
depending on the set of SUSY parameters.
The overall magnitude of the $\bar \sigma_0, ~\bar \sigma_{22}$
cross section in the $A^0$ case tend to be considerably larger
than those of the $H^0$ one.

\section{Conclusions}

In this paper we have discussed the properties of the
process $\gamma\gamma\to ZH$ where $H$ is either the SM Higgs boson
$H_{SM}$, or any one of the three neutral supersymmetric Higgs bosons
$h^0, ~H^0, ~A^0$. These processes only arise at the 1-loop level, involving
triangle $H-\gamma\gamma$ and $Z,G^0-\gamma\gamma$
diagrams, as well as $\gamma\gamma ZH$-box
diagrams with internal charged fermionic lines ($l,q,\chi^{\pm}_i$).
We have shown how these contributions reflect in the
$\gamma\gamma\to ZH$ observables.\par

It appears that for all 4 cases, the helicity properties of the
amplitudes are very simple.
The final $Z$ is almost always in the longitudinal state;
\ie for more than $98\%$ of the cases. Moreover,
all these processes  occur for more than $95\%$ of the times for
initial photon-photon helicities in the $\Delta\lambda=0$ configuration.
This implies that there is  essentially  only one amplitude
 contributing; namely the $F_{\pm\pm0}$ leading to
\bq
\sigma_0\simeq\sigma_{22}\simeq\eta\sigma'_{33} ~~ ,
\eq
with $\eta=-1$ for $(H_{SM}, ~h^0, ~H^0)$, and   $\eta=+1$ for $A^0$.

The $ZH$ angular distribution is always rather flat, so that an
important part of the events are produced at large angles,
facilitating the detection.

The most spectacular properties concern the energy dependence of the
cross section, which show strong threshold effects, due mainly
to the fermionic box amplitudes. They are induced by
the standard top quark
and the supersymmetric $\chi^{\pm}_i$ chargino contributions.
Depending on
the type of the neutral Higgs boson produced  and on
the domain of MSSM parameter space, one can observe well pronounced
threshold effects with steps, bumps or peaks. We have given
typical illustrations in Figs.3-6 using two rather extreme  sets of SUSY
parameters.

We  conclude by emphasizing that the neutral Higgs production
processes considered here, provide
remarkable  tests of the structure of the electroweak
interactions, which are complementary to those encountered
in the gauge sector through studies of the
$\gamma\gamma \to \gamma\gamma ,~ \gamma Z, ~ZZ$ transitions;
and to the tests of the Higgs sector provided by $\gamma\gamma \to H$.

Although the cross sections seem rather small, several effects
appear to be very spectacular. It appears therefore worthwhile
that these processes are considered by the working groups, in order to
study their observability at future high energy and
high luminosity photon-photon colliders.

\vspace{1cm}
\noindent
{\bf Acknowledgments.}\\
 We are pleased to thank Abdelhak  Djouadi for very
informative discussions.

\newpage

\renewcommand{\theequation}{A.\arabic{equation}}
\renewcommand{\thesection}{A.\arabic{section}}
\setcounter{equation}{0}
\setcounter{section}{0}

{\large \bf Appendix A:  The needed couplings in
 the Standard and SUSY models.}

We generally give the couplings  in SUSY models,
specifying also the limit at which the SM ones are recovered.
We use the same notation as in the Appendix of \cite{ggZZ-2nd},
giving here  only  the couplings needed in
the present calculation.
These consist of the photon- and
Z-fermion ones  determined by
\bqa
\L_{Vff} & =& -e Q_f A^\mu \bar f\gamma_\mu f  - e Z^\mu \bar f
(\gamma_\mu g_{vf}^Z- \gamma_\mu \gamma_5 g_{af}^Z) f ~~ ,
\nonumber \\
 &  & -e A^\mu \bar{\tchi}_j \gamma_\mu \tchi_j
-e   Z^\mu \bar{\tchi}_j \left  ( \gamma_\mu g_{vj}^Z
- \gamma_\mu \gamma_5 g_{aj}^Z \right ) \tchi_j
\nonumber \\
&& -e  Z^\mu \left [\bar{\tchi}_1 \left
( \gamma_\mu g_{v12}^Z
- \gamma_\mu \gamma_5 g_{a12}^Z \right ) \tchi_2 +
\mbox{h.c.} \right ]
  ~~ , \label{gauge-fermion-vertex}
\eqa
where $f$ is an ordinary quark or lepton  and $\tchi_j ~(j=1,2)$
are  the two
positively charged  charginos. From this we have
\bq
g_{vf}^Z=\frac{t_3^f-2Q_f\sw^2}{2\sw\cw} ~~~~~ ,
~~~~~ g_{af}^Z=\frac{t_3^f}{2\sw\cw} ~ ~~~~
\label{Z-f-coupling}
\eq
for the $Zff$-couplings, while Z-charginos ones are written
as
\bqa
g_{v1}^Z&= &\frac{1}{2\sw\cw}\left (
{3\over2}-2s^2_W+{1\over4}[\cos2\phi_L+\cos2\phi_R]\right ) ~ ,
 \nonumber \\
g_{a1}^Z& = & - ~\frac{1}{8\sw\cw}[\cos2\phi_L-\cos2\phi_R] ~ ,
\label{Z-chi1-coupling} \\
g_{v2}^Z& = & \frac{1}{2\sw\cw}\left ({3\over2}-2s^2_W-~
{1\over4}[\cos2\phi_L+\cos2\phi_R] \right )
~ , \nonumber \\
g_{a2}^Z& = & \frac{1}{8\sw\cw}[\cos2\phi_L-\cos2\phi_R] ~ ,
\label{Z-chi2-coupling} \\
g_{v12}^Z& = & g_{v21}^Z=
-~{\Sn(M_2)\over 8\sw\cw }
[\B_R ~ \Del_{12} \sin2\phi_R+ \B_L \sin2\phi_L] ~ ,
\nonumber\\
g_{a12}^Z&= & g_{a21}^Z= -~{\Sn(M_2)\over 8\sw\cw}
[\B_R ~  \Del_{12}\sin2\phi_R - \B_L \sin2\phi_L] ~ .
\label{Z-chi12-coupling}
\eqa
The sign quantities $(\Del_{12}, ~\B_L,
~\B_R)$ in (\ref{Z-chi12-coupling}) are
 related  to the  definition of the chargino mixing angles,
which is selected to always  obey $0\leq \phi_L ,
~\phi_R\leq\pi/2$. They are given in Eqs.(A.35) in the Appendix of
\cite{ggZZ-2nd}.\par

Also needed are the Yukawa couplings of the neutral Higgs bosons
to the ordinary fermions and charginos determined by the effective
Lagrangian
\bqa
&& \L_{\rm Yukawa}=(g_{H^0ff} H^0 +g_{h^0ff} h^0)\bar f f +
i \tilde g_{A^0ff} A^0 \bar f \gamma_5 f+
(g_j^{h^0} h^0 +g_j^{H^0} H^0) \bar{\tchi}_j \tchi_j
\nonumber \\
&& + i \tilde g_j^{A^0} A^0 \bar{\tchi}_j \gamma_5 \tchi_j
+ \big [(g_{s12}^{h^0} h^0 + g_{s12}^{H^0} H^0 +
g_{s12}^{A^0} A^0)\bar{\tchi}_1 \tchi_2
\nonumber \\
&& + (g_{p12}^{h^0} h^0 + g_{p12}^{H^0} H^0 +
g_{p12}^{A^0} A^0)\bar{\tchi}_1 \gamma_5 \tchi_2 +
{\rm h.c.} \big ] ~~, \label{Yukawa-vertex}
\eqa
which for the  quarks and leptons of the
third family (the only ones needed to be retained) give
\bqa
&& g_{h^0tt}=-\frac{g \mt}{2 \mw} \frac{\calpha}{\sbeta} ~~,~~
g_{H^0tt}=-\frac{g \mt}{2 \mw} \frac{\salpha}{\sbeta}~~,~~
\tilde g_{A^0tt}= \frac{g \mt}{2 \mw} \cot\beta
\nonumber \\
&& g_{h^0bb}= \frac{g m_b}{2 \mw} \frac{\salpha}{\cbeta} ~~,~~
g_{H^0bb}=-\frac{g m_b}{2 \mw} \frac{\calpha}{\cbeta}~~,~~
\tilde g_{A^0bb}= \frac{g m_b}{2 \mw} \tan \beta
\nonumber  \\
&& g_{h^0\tau \tau}= \frac{g m_\tau}{2 \mw} \frac{\salpha}{\cbeta} ~~,~~
g_{H^0\tau\tau}=-\frac{g m_\tau}{2 \mw} \frac{\calpha}{\cbeta}~~,~~
\tilde g_{A^0\tau\tau}= \frac{g m_\tau}{2 \mw} \tan \beta ~.
\label{H-f-coupling}
\eqa\par

Parameters  $\alpha, ~\beta$ are the usual SUSY Higgs sector angles.
In the SM case, the couplings of $H_{\rm SM}$ should be identified
with those of $h^0$ by
putting $\alpha=\beta -\pi/2$.
Finally the Higgs-chargino couplings in (\ref{Yukawa-vertex}) are
given by
\bqa
g_1^{h^0} &=& -\frac{g}{\sqrt 2}\Del_1
( -\cfR \sfL \salpha ~\B_L +\sfR \cfL \calpha ~\B_R) ~,
\nonumber \\
g_1^{H^0} &=& -\frac{g}{\sqrt 2}\Del_1
( \cfR \sfL \calpha ~\B_L +\sfR \cfL \salpha ~\B_R) ~,
\nonumber \\
\tilde g_1^{A^0} &=& -\frac{g}{\sqrt 2}\Del_1
( \cfR \sfL \sbeta ~\B_L +\sfR \cfL \cbeta ~\B_R)  ~,
\label{H-chi1-coupling} \\
g_2^{h^0} &=& -\frac{g}{\sqrt 2}\Del_2
(- \cfR \sfL \calpha ~\B_L + \sfR \cfL \salpha ~\B_R )  ~,
\nonumber \\
g_2^{H^0} &=& \frac{g}{\sqrt 2}\Del_2
( \cfR \sfL \salpha ~\B_L +\sfR \cfL \calpha ~\B_R) ~,
\nonumber \\
\tilde g_2^{A^0} &=& \frac{g}{\sqrt 2}\Del_2
( \cfR \sfL \cbeta ~\B_L +\sfR \cfL \sbeta ~\B_R)  ~,
\label{H-chi2-coupling}
\eqa
for the lighter and heavier chargino denoted
as $\tchi_1$ and $\tchi_2$ respectively. As in the case of
(\ref{Z-chi12-coupling}), the
sign-quantities $\Del_1, ~ \Del_2, ~\B_L, ~\B_R$ are also
related to the
chargino mixing and defined in (A.35) of the Appendix of \cite{ggZZ-2nd}.
Finally  the mixed  Higgs-chargino couplings are
\bqa
&& g^{h^0}_{s12}=g^{h^0}_{s21} =
\frac{g}{2\sqrt{2}} \Sn (M_2) (\Del_1\calpha-\Del_2\salpha)
[\B_{LR}\sfL \sfR-\Del_{12} \cfL\cfR  ] ,
\nonumber \\
&& g^{h^0}_{p12}=- g^{h^0}_{p21} =
\frac{g}{2\sqrt{2}} \Sn (M_2) (\Del_1\calpha +\Del_2\salpha)
[\B_{LR}\sfL \sfR +\Del_{12} \cfL\cfR  ] ,
\nonumber \\
&& g^{H^0}_{s12}=g^{H^0}_{s21} =
\frac{g}{2\sqrt{2}} \Sn (M_2) (\Del_1\calpha +\Del_2\salpha)
[-\cfL \cfR+ \Del_{12}\B_{LR} \sfL\sfR ] ,
\nonumber \\
&& g^{H^0}_{p12}=- g^{H^0}_{p21} = -
\frac{g}{2\sqrt{2}} \Sn (M_2) (\Del_1\calpha -\Del_2\salpha)
[\cfL \cfR +\Del_{12}\B_{LR} \sfL\sfR  ] ,
\nonumber \\
&& g^{A^0}_{s12}= - g^{A^0}_{s21}
\nonumber \\
&& =-i \frac{g}{2\sqrt{2}} \Sn (M_2) (\Del_1\sbeta -\Del_2\cbeta)
[\cfL \cfR+ \Del_{12}\B_{LR} \sfL\sfR ] ,
\nonumber \\
&& g^{A^0}_{p12}=  g^{A^0}_{p21}
\nonumber \\
&& =-i \frac{g}{2\sqrt{2}} \Sn (M_2) (\Del_1\sbeta + \Del_2\cbeta)
[\cfL \cfR - \Del_{12}\B_{LR} \sfL\sfR ] .
\label{H-chi12-coupling}
\eqa

\vspace{1cm}

\renewcommand{\theequation}{B.\arabic{equation}}
\renewcommand{\thesection}{B.\arabic{section}}
\setcounter{equation}{0}
\setcounter{section}{0}

{\large \bf Appendix B:  The $\gamma \gamma \to Z h^0, ~ ZH^0,~ZH_{SM}$
helicity amplitudes.}

The invariant helicity amplitudes for  the process
$\gamma \gamma \to Z h^0$, (or $Z H^0$ or $ZH_{SM}$)
\bq
\gamma (k_1,\lambda_1) \gamma (k_2,\lambda_2) \to
Z (q_1,\lambda_Z) ~h^0 (q_2) \ \ ,
\label{ggZh0-process}
\eq
are denoted as
$F_{\lambda_1 \lambda_2 \lambda_Z}(\kappa ,t,u)$,
where the momenta and
helicities of the incoming  photons and outgoing $Z$'s
 are indicated in parentheses, and
\bqa
&& s=(k_1+k_2)^2 ~~ ~,~ ~ t=(k_1-q_1)^2 ~ ~,~ ~
u=(k_1-q_2)^2  ~ , \nonumber \\
&& \kappa=[s-(m-\mz)^2]^{1/2} [s-(m+\mz)^2]^{1/2} ~ . \label{kin1}
\eqa
Here  $m$ stands for the mass of the neutral Higgs boson
in the final state. In the present case this is the
mass of  $h^0$, (or $H^0,~ H_{SM}$),
but similar definitions will also be used  for the  $A^0$ production
case.  Notice also that  in the  $\gamma \gamma $ c.m. frame
\bq
|\vec q_1|=|\vec q_2|=\frac{\kappa}{2\sqrt{s}} ~ . \label{kin2}
\eq \par

The number of independent helicity amplitudes is reduced by
various symmetries. Thus, if
the only existing CP-violation is the usual one
related to the standard part of the Yukawa forces;
then at the 1-loop level the amplitudes should be CP invariant
implying
\bq
F_{\lambda_1, \lambda_2, \lambda_Z}(\kappa ,t,u)=-
F_{-\lambda_1, -\lambda_2, -\lambda_Z}(\kappa
,t,u)(-1)^{\lambda_Z} ~ ,\label{CP-h0-con}
\eq
while Bose statistics imposes
\bq
F_{\lambda_1 \lambda_2 \lambda_Z}(\kappa ,t,u)=
F_{\lambda_2 \lambda_1 \lambda_Z}(\kappa ,u,t)(-1)^{\lambda_Z} ~,
\label{Bose-h0-con}
\eq
and  the standard properties of the Z-polarization vectors
give \cite{ggZZ}
\bq
F_{\lambda_1, \lambda_2, \lambda_Z}(\kappa ,t,u)=-
F_{\lambda_1, \lambda_2, -\lambda_Z}(-\kappa
,t,u)(-1)^{\lambda_Z} ~ .\label{Zpol-h0-con}
\eq

Therefore,  there are only four independent helicity amplitudes
  which are  taken as
\bq
F_{+++}~~~,~~~ F_{+--}~~~,~~~ F_{++0}~~~,~~~ F_{+-0} ~~~
\label{basic-h0-amplitudes}
\eq
and referred to below as "basic" amplitudes. The other
amplitudes  are determined by
\bqa
F_{++-}(\kappa , t,u ) &= & F_{+++}(-\kappa, t,u) ~~,
\nonumber \\
F_{+-+}(\kappa, t, u)&=& F_{+--}(-\kappa, t,u) ~~,
\label{ampl-con1}
\eqa
and  (\ref{CP-h0-con}). On the
basis of (\ref{CP-h0-con}-\ref{Zpol-h0-con}), we also note that
\bqa
&& F_{++0}(\kappa , t,u ) =  - F_{++0}(-\kappa, t,u) ~~,
\label{ampl-con2} \\
&& F_{+-0}(\kappa, t, u)=- F_{+-0}(-\kappa, t, u)=
 -F_{-+0}(\kappa, t, u)
\nonumber \\
&& =- F_{+-0}(\kappa, u, t)= F_{-+0}(\kappa, u, t) ~~ .
\label{ampl-con3}
\eqa\par

At the 1-loop level, these amplitudes are expressed  in terms
of the $C_0$ and $D_0$ Passarino-Veltman functions
\cite{Passarino}, for which we follow the notation of
\cite{Hagiwara} and the abbreviations\footnote{In
(\ref{C0sabc}-\ref{DhZtuabba}), the momenta $p_1=k_1$, $p_2=k_2$ denoting
the momenta of the photons, and   $p_3=-q_1$, $p_4=-q_2$
being opposite to those of the final Z and  $h^0$, are
always taken as incoming; compare (\ref{ggZh0-process}).}
\bqa
&& C_{0}^{abc}(s)\equiv
 C_0(p_1, p_2; m_{a},m_{b},m_{c}) =  C_{0}(0,0,s; m_{a},m_{b},m_{c}) ~ ,
\label{C0sabc} \\
&& C_{h}^{abc}(u)\equiv
C_0(p_4, p_1; m_{a},m_{b},m_{c}) =  C_{0}(m^2,0,u; m_{a},m_{b},m_{c})
~ ,\label{Chuabc} \\
&& C_{Z}^{abc}(u)\equiv
C_0(p_3, p_2; m_{a},m_{b},m_{c}) =  C_{0}(m_Z^2,0,u; m_{a},m_{b},m_{c})
~ ,\label{CZuabc} \\
&& C_{h}^{abc}(t)\equiv
C_0(p_4, p_2; m_{a},m_{b},m_{c}) =  C_{0}(m^2,0,t; m_{a},m_{b},m_{c})
~ ,\label{Chtabc} \\
&& C_{Z}^{abc}(t)\equiv
C_0(p_3, p_1; m_{a},m_{b},m_{c}) =  C_{0}(m_Z^2,0,t; m_{a},m_{b},m_{c})
~ ,\label{CZtabc} \\
&& C_{hZ}^{abc}(s) \equiv C_0(p_4, p_3; m_{a},m_{b},m_{c})  =
C_{0}(m^2,m_{Z}^2,s;m_{a},m_{b},m_{c})
~ . \label{ChZsabc}
\eqa \par

Correspondingly for the $D_0$-functions, we  note  that
\bqa
&& D_{hZ}^{abcd}(s,u) \equiv D_0(p_4, p_3, p_2; m_a,m_b,m_c,m_d) =
\nonumber \\
&& D_{0}(m^2,m_{Z}^2,0,0,s,u;m_a,m_b,m_c,m_d)
~ , \label{DhZsuabcd}  \\
&& D_{hZ}^{abcd}(s,t) \equiv D_0(p_4, p_3, p_1; m_a,m_b,m_c,m_d) =
\nonumber \\
&& D_{0}(m^2,m_{Z}^2,0,0,s,t ;m_a,m_b,m_c,m_d)
~ , \label{DhZstabcd}
\eqa
which for a common propagator mass simplify to
\bqa
&& D_{hZ}^{f}(t,u)\equiv D_0(p_4, p_2, p_3; m_f)
 = D_{0}(m^2,0,m_Z^2,0,t,u; m_f,m_f,m_f,m_f)
\nonumber \\
&& =D_0(p_4, p_1, p_3; m_f)=D_{0}(m^2,0,m_Z^2,0,u,t; m_f,m_f,m_f,m_f)
\nonumber \\
&& = D_{0}(p_3,p_2,p_4;m_f)=
D_{0}(m_Z^2,0,m^2,0,u,t; m_f,m_f,m_f,m_f) . \label{DhZtuf}
\eqa
In the same spirit, when \eg  $m_a=m_b=m_c$~,
the Passarino-Veltman C-functions
 are further abbreviated like in $C_{0}^{abc}(s)\Rightarrow C_0^a(s)$.

Correspondingly,  for the case of  two different
propagator masses in a D-function we have
\bqa
&& D_{hZ}^{abba}(t,u)\equiv D_0(p_4, p_2, p_3; m_a,m_b,m_b, m_a)
 = D_0(p_4, p_1, p_3; m_b,m_a,m_a, m_b)
\nonumber \\
&& =D_0(p_3, p_2, p_4; m_a,m_b,m_b, m_a)
 = D_0(p_3, p_1, p_4; m_b,m_a,m_a, m_b)
\nonumber \\
&& =D_0(m^2,0,m_Z^2,0,t,u; m_a,m_b,m_b,m_a)
=D_{0}(m^2,0,m_Z^2,0,u,t; m_b,m_a,m_a,m_b)
\nonumber \\
&& =D_0(m_Z^2,0,m^2,0,u,t; m_a,m_b,m_b,m_a)
\nonumber \\
&& =D_{0}(m_Z^2,0,m^2,0,t,u; m_b,m_a,m_a,m_b)~ . \label{DhZtuabba}
\eqa
Notice that (\ref{DhZtuabba}) imply that
\bq
D_{hZ}^{abba}(t,u)=D_{hZ}^{baab}(u,t) ~~.
\eq\par

For the charginos boxes  below, instead of the notation \eg
$C_0^{\tchi_2 \tchi_1 \tchi_1}(s)$, we write $C_0^{211}(s)$. \par

As in \cite{ggZZ, ggAA, ggZZ-2nd},  it is convenient to
define
\bqa
&& Y=t u -m^2 \mzd ~~ ,~~ s_h=s-m^2 ~~,~~ t_h=t-m^2~~,~~
 u_h=u-m^2 ~~,\nonumber \\
&& s_Z=s-\mzd ~~,~~  u_Z=u-\mzd ~~~ , ~~~ t_Z=t-\mzd ~~~,~~~
\label{kin3} \\
&& \tF^f(s,t,u)=D^f_{hZ}(s,u)+D^f_{hZ}(s.t)+D^f_{hZ}(t,u) ~~ ,
\nonumber \\
&&E_1^f(s,u)=u_h C_h^f(u)+u_Z C_Z^f(u)-s u D_{hZ}^f(s,u) ~~,
\nonumber \\
&& E_2^f(t,u)=u_h C_h^f(u)+u_Z C_Z^f(u)+t_h C_h^f(t)+t_Z C_Z^f(t)
- YD_{hZ}^f(t,u) ~~ , \label{new-functions-1}
\nonumber \\
&& E_1^{ab}(s,u)=u_h C_h^{baa}(u)+u_Z C_Z^{baa}(u)
-s u D_{hZ}^{abaa}(s,u) ~~,
\nonumber \\
&& E_2^{ab}(t,u)=u_h C_h^{baa}(u)+u_Z C_Z^{baa}(u)+t_h C_h^{baa}(t)
\nonumber \\
&& +t_Z C_Z^{baa}(t) - YD_{hZ}^{abba}(t,u) ~~ . \label{new-functions-2}
\eqa
Notice that $\tF^f(s,t,u)$, $E_2^f(t,u)$ and  $E_1^{ab}(s,u)$,
$E_1^f(s,u)$
remain the same under interchanging $m^2 \leftrightarrow \mzd $;
while  $E_2^{ab}(t,u)$ remains the same under
($m^2 \leftrightarrow \mzd $ and $t \leftrightarrow u $ ).\par

\begin{figure}[p]
\vspace*{-2cm}
\[
\epsfig{file=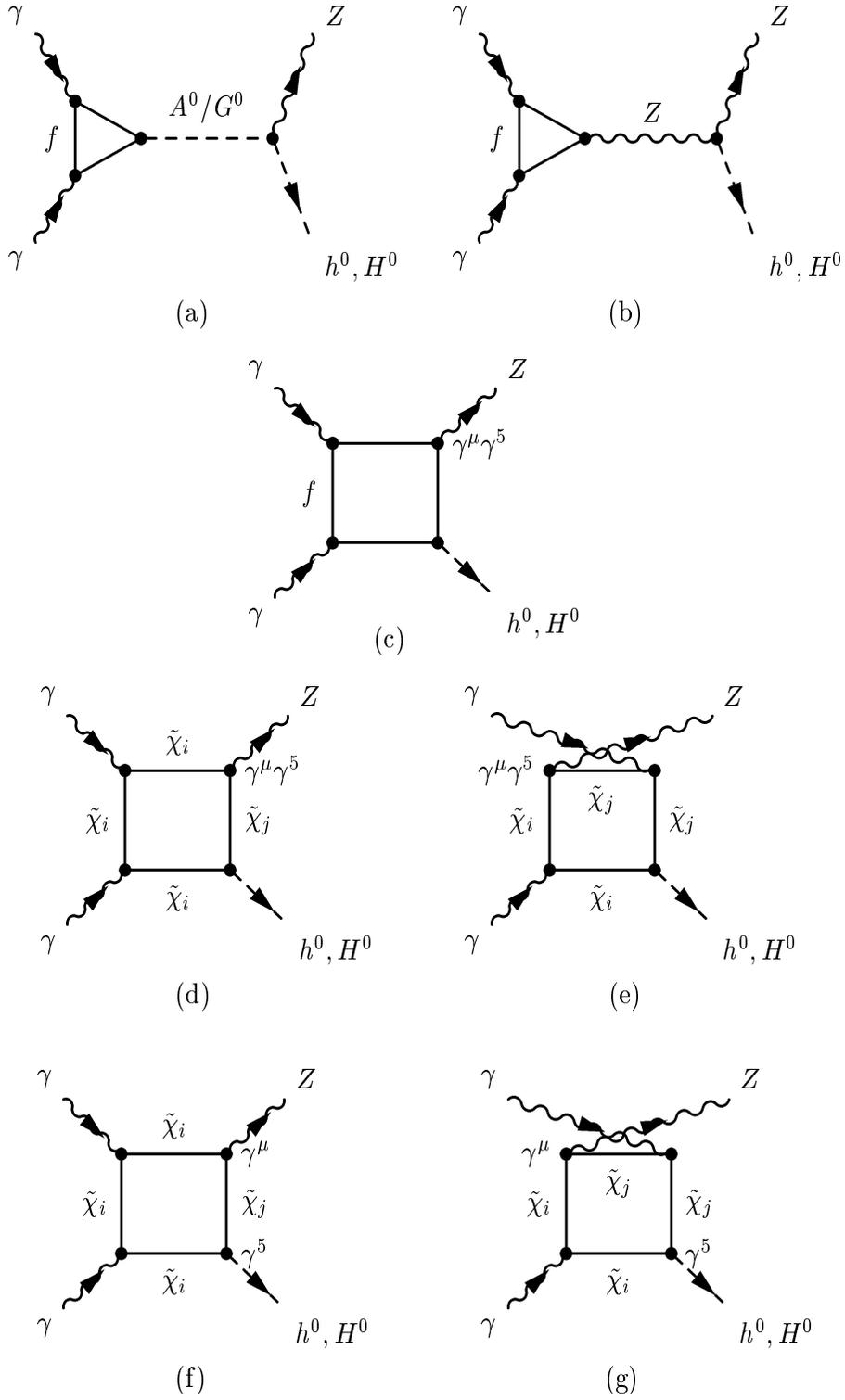,height=20cm,width=12cm}
\]
\caption[1]{Generic diagrams describing the various
contributions to $\gamma \gamma \to Z h^0, ~Z H^0$ in SUSY models.
 Solid lines correspond to fermions, broken lines  to scalars, while
wavy lines correspond to gauge bosons. Similar diagrams also
describe the Standard Model. The diagrams
in (d-g) for $j \neq i$ describe the mixed chargino boxes. }
\label{ggZh0-diag}
\end{figure}

\vspace{1cm}
{\bf \underline{ The  $A^0$-pole contribution.}} It only exists
in SUSY models and it is  described by the diagram in
Fig.\ref{ggZh0-diag}a, in which only $A^0$ exchange is considered.
The fermion loop determining the $\gamma \gamma A^0$ vertex of
this diagram involves essentially
only the $t$ and $b$ quarks, the $\tau$-leptons
and the  charginos. The only non-vanishing contribution from each
of these fermions to the basic amplitudes
appearing in (\ref{basic-h0-amplitudes}), is
for\footnote{Notice that  $\alpha$ is used to describe both the fine
structure constant, as well as the usual Higgs sector mixing angle.
The discrimination among them in each case,
should be easy  though, from the structure of the formulae.}
\bq
F_{++0}^{A0f-\rm pole}(\gamma \gamma \to Z h^0)
=-\frac{\alpha g Q_f^2 N_f^c}{2\pi \mw}~ \frac{\tilde
g_{A^0ff}\cos(\alpha-\beta)}{s-m_{A^0}^2+im_{A^0} \Gamma_{A^0}}
~ \kappa m_f s C_0^f(s) ~~, \label{A0-pole-f-h0}
\eq
where for quarks and leptons of the third family
$\tilde g_{A^0ff}$ is given in
(\ref{H-f-coupling}); while for the two  charginos
the corresponding couplings are given by $g_1^{A^0}$ and $g_2^{A^0}$
in (\ref{H-chi1-coupling}), (\ref{H-chi2-coupling}) respectively.
In (\ref{A0-pole-f-h0}) $N_f^c$ is the colour factor, being 3 for
quarks, and 1 for $\tau$'s and the charginos. As usually
$g=e/sw$. \par

The corresponding contribution to the $\gamma \gamma \to Z H^0$
process is given from (\ref{A0-pole-f-h0}) by replacing
\bq
\cos(\alpha-\beta) \Rightarrow -\sin(\beta-\alpha) ~~ .
\label{A0-pole-f-H0}
\eq\par

\vspace{1cm}
{\bf \underline{ The  $Z-G^0$-exchange contribution.}} It is
described  by the diagram in Fig.\ref{ggZh0-diag}b for the
Z-exchange part, together with the neutral Goldstone exchange
indicated in Fig.\ref{ggZh0-diag}a. In both cases the physical
contribution only arises from the  spin=0 part of the propagator
exchanged in the s-channel, and there is no pole at $\mzd$.
Notice that the diagram Fig.\ref{ggZh0-diag}b
would also create a
$Z\gamma\gamma$ anomaly, which is being of course cancelled when
 a complete family of quarks and leptons or both charginos are
included. The only non-vanishing contributions from these
diagrams to the  basic amplitudes of  (\ref{basic-h0-amplitudes})
are
\bq
F^{tb\tau Z-\rm pole}_{++0}(\gamma \gamma \to Z h^0)= -
\frac{\kappa \alpha^2\sin(\beta-\alpha)}{\swd \cwd \mzd}
\Big [\frac{4\mt^2}{3} C_0^t(s)-\frac{m_b^2}{3}C_0^b(s)-
m_\tau^2 C_0^\tau (s) \Big ] ~~ \label{Z-pole-tbtau-h0}
\eq
due to the mass differences among
 the quarks and leptons of the third
family\footnote{The contributions from the first
two families is negligible due to their small masses.}, and
\bqa
F^{\tchi_1 \tchi_2 Z-\rm pole}_{++0}(\gamma \gamma \to Z h^0)&=&
\frac{\kappa \alpha^2\sin(\beta-\alpha)}{2 \swd \cwd \mzd}
[\cos(2\phi_L)-\cos(2\phi_R)]
\nonumber \\
&& \cdot \Big [m_{\tchi_1}^2C_0^{\tchi_1}(s)-
m_{\tchi_2}^2C_0^{\tchi_2}(s) \Big ] ~~ \label{Z-pole-chi12-h0}
\eqa
from the two charginos.\par

The corresponding contribution to the $\gamma \gamma \to Z H^0$
process is given from (\ref{Z-pole-tbtau-h0}, \ref{Z-pole-chi12-h0})
by replacing
\bq
\sin(\beta-\alpha) \Rightarrow  \cos(\beta-\alpha) ~~ .
\label{Z-pole-h0-H0}
\eq

\vspace{1cm}
{\bf \underline{Single fermion box contribution.}}
The generic single fermion $f$-box diagram
inducing this contribution, is shown in
Fig.\ref{ggZh0-diag}c, where only the axial part of $Z$ contributes.
We write this contribution  as
\bq
F^{f-\rm box}_{\lambda_1\lambda_2\lambda_Z}(\gamma \gamma \to Z
h^0 ~(H^0))= r_f^{h^0 (H^0)}\cdot
~ A^{f-\rm box}_{\lambda_1\lambda_2\lambda_Z}(H) ~~ .
\label{f-box-amplitude-h0}
\eq
The relevant couplings are collected in the
coefficients, which for quarks or leptons are written as
\bqa
r_f^{h^0} &=& \frac{e^3}{(4\pi)^2} N_f^c Q_f^2 g_{af}^Z g_{h^0ff}
~ ~,
\nonumber \\
r_f^{H^0} &=& \frac{e^3}{(4\pi)^2} N_f^c Q_f^2 g_{af}^Z g_{H^0ff}
~~ , \label{r-f-box-H}
\eqa
(compare (\ref{Z-f-coupling}, \ref{H-f-coupling}).
The same  expression
also applies to the standard $H_{SM}$ production process.
Correspondingly, for a  box with  single chargino
running along its sides,
 we have
\bq
r_f^{h^0(H^0)}=\frac{e^3}{(4\pi)^2} g_{aj}^Z g_j^{h^0 (H^0)}
~~  \label{r-chi-box-H}
\eq
with the couplings given in (\ref{Z-chi1-coupling},
\ref{Z-chi2-coupling}, \ref{H-chi1-coupling},
\ref{H-chi2-coupling}). \par

The $A^{f-\rm box}_{\lambda_1\lambda_2\lambda_Z}(H)$ terms  in
(\ref{f-box-amplitude-h0}) are then given by
\bqa
&& A^{f-\rm box}_{+++}(H)=- \frac{\sqrt{2} m_f}{\kappa \sqrt{Ys}}
\Bigg \{ s(t-u)(t_h+u_h-\kappa)C_0^f(s)
+2u_h[t_h(t-\kappa)+m^2 u_Z+Y]C_h^f(u)
\nonumber \\
&&-2u_Z[u_h(u-\kappa)+m^2 t_Z+Y]C_Z^f(u)
+s(t_h+u_h-\kappa)(Y+u^2-m^2\mzd )D_{hZ}^f(s,u)
\nonumber \\
&& -\frac{Y}{2} (t-u)(t_h+u_h-\kappa)D_{hZ}^f(t,u)- ~
(t \leftrightarrow u ) \Bigg \} ~~ , \label{h0-f-box+++}
\\
&& A^{f-\rm box}_{+--}(H)= - \frac{\sqrt{2}m_f}{\kappa
\sqrt{s Y^3}}(u-t+\kappa)\Bigg \{
s\Big [ \kappa (t_h t_Z+ u_h u_Z+Y)+s (u^2-t^2)\Big ]C_0^f(s)
\nonumber \\
&& +\kappa (Y+2 u_hu_Z)[u_h C_h^f(u)+u_Z C_Z^f(u)]
+Y(u_Z+t_Z)[u_Z C_Z^f(u)-u_h C_h^f(u)]
\nonumber \\
&& + s\kappa [t^2+u^2- 2m^2 \mzd +(t-u)\kappa ]C_{hZ}^f(s)
+2 s (u^2-m^2 \mzd)E_1^f(s,u)
\nonumber \\
&& +2 s m_f^2 Y (\kappa +t-u) \tF^f(s,t,u)
-\kappa s u [2 u (u_h+t_Z)-Y]D_{hZ}^f(s,u)
\nonumber \\
&& +\frac{Y^2 \kappa}{2} D_{hZ}^f(t,u)- ~
(t \leftrightarrow u ~,~ \kappa \rightarrow -\kappa) \Bigg \}
~~ ,\label{h0-f-box+--}
\\
&& A^{f-\rm box}_{++0}(H)= \frac{4 m_f}{\kappa \mz s} \Bigg \{
2 s^2 (t+u)C_0^f(s)+[(t+u)(m^2+\mzd)-4 m^2\mzd]E_2^f(t,u)
\nonumber \\
&& -2 m_f^2  s \kappa^2\tF^f(s,t,u)-2 m^2 \mzd s^2
[D_{hZ}^f(s,u)+D_{hZ}^f(s,t)] \Bigg \} ~~ ,
\label{h0-f-box++0} \\
&& A^{f-\rm box}_{+-0}(H)=\frac{4 m_f}{\kappa \mz Y} \Bigg \{
(t-u)(t_Z+u_Z)\Big [s (t+u)C_0^f(s)-\kappa^2 C_{hZ}^f(s) -2 m_f^2 Y
\tF^f(s,t,u)\Big ]
\nonumber \\
&& + (t_Z+u_Z)\Big [ (t^2-m^2 \mzd)E_1^f(s,t)-
(u^2-m^2 \mzd)E_1^f(s,u) \Big ]
\nonumber \\
&& + 2\mzd Y\Big [u_h C_h^f(u)-t_h C_h^f(t)-u_Z C_Z^f(u)+t_Z C_Z^f(t)
\Big ] \Bigg \} ~~ . \label{h0-f-box+-0}
\eqa\par

\vspace{1cm}
{\bf \underline{Mixed chargino box involving axial Z coupling.}}
The generic form of the box diagrams giving this
contribution is shown in Fig.\ref{ggZh0-diag}d,e. Their
characteristic feature is that they involve the mixed
 axial Z-coupling of (\ref{Z-chi12-coupling}) and
the $g_{s12}^{h^0}$, $g_{s12}^{H^0}$ type of Higgs
couplings appearing in (\ref{H-chi12-coupling}). Notice that the
diagrams of type (d) involve three identical chargino masses of one kind,
and   one of the other. On the contrary, the diagram of
type (e) has two $\tchi_1$-propagators  and two of $\tchi_2$.
In analogy to (\ref{f-box-amplitude-h0}), their contribution is
written  as
\bq
F^{Z_a \tchi_1\tchi_2 -\rm box}_{\lambda_1\lambda_2\lambda_Z}
(\gamma \gamma \to Z h^0 (H^0) )=
r_{Z_a \chi_1\chi_2}^{h^0 (H^0)} \cdot
A^{Z_a \tchi_1\tchi_2 -\rm box}_{\lambda_1\lambda_2\lambda_Z}(H) ~~ ,
\label{Zachi12-box-amplitude-h0}
\eq
where the various couplings are absorbed in the coefficients
\bqa
r_{Z_a \chi_1\chi_2}^{h^0}&=&
\frac{e^3}{(4\pi)^2} g_{a12}^Z g_{s12}^{h^0} ~~ ,
\nonumber \\
r_{Z_a \chi_1\chi_2}^{H^0}&=&
\frac{e^3}{(4\pi)^2} g_{a12}^Z g_{s12}^{H^0} ~~ ,
\label{r-Za-chi12-box-H}
\eqa
for the $h^0$ and $H^0$ production respectively.\par

The $A^{Z_a \tchi_1\tchi_2 -\rm box}_{\lambda_1\lambda_2\lambda_Z}$
terms  in (\ref{Zachi12-box-amplitude-h0}) are then given by
\bqa
&& A^{Z_a \tchi_1\tchi_2 -\rm box}_{+++}(H) =
- \frac{\sqrt{2}}{\kappa \sqrt{s Y}} \Bigg \{
\Big \{ s \mchil (t-u) (t_h+u_h-\kappa) C_0^{111}(s)
\nonumber \\
&& +(\mchil+\mchih)\Big [u_h [t_h(t-\kappa) +m^2 u_Z+Y]C_h^{211}(u)
 -u_Z [u_h(u-\kappa) +m^2 t_Z+Y]C_Z^{211}(u) \Big ]
\nonumber \\
&& + s \mchil (t_h+u_h-\kappa)
[Y+ u^2 -m^2 \mzd -(\mchil^2-\mchih^2)(t-u)]D_{hZ}^{1211}(s,u)
\nonumber \\
&& - \frac{(t-u)}{8}(\mchil+\mchih)(t_h+u_h -\kappa)[s (\mchil-\mchih)^2+Y]
[D_{hZ}^{1221}(t,u)+D_{hZ}^{2112}(t,u)]
\nonumber \\
&& -\frac{\kappa (\mchil-\mchih)}{8}(t_h+u_h-\kappa)[s(\mchil+\mchih)^2+Y]
[D_{hZ}^{1221}(t,u)-D_{hZ}^{2112}(t,u)]
\nonumber \\
&& - (t \leftrightarrow u) \Big \} + ( 1 \leftrightarrow  2) \Bigg \}
~~ , \label{h0-Zachi12-box+++} \\
&& A^{Z_a \tchi_1\tchi_2 -\rm box}_{+--}(H)= -\frac{(\kappa
-t+u)}{\kappa \sqrt{2 s Y^3}} \Bigg \{ \Bigg [
s\Big \{ s (\mchil+\mchih)(u-t)[u+t+2(\mchil^2-\mchih^2)]
\nonumber \\
&& + \kappa [\mchil (t_Z t_h+u_Zu_h)-\mchih s
(t+u)-2s(\mchil^2-\mchih^2)(\mchil+\mchih)]\Big \} C_0^{111}(s)
\nonumber \\
&& +(\mchil+\mchih)\Big \{
[2 s (u^2-m^2\mzd )+\kappa (Y+2 u_hu_Z)]
[u_ZC_Z^{211}(u)+u_hC_h^{211}(u)]
\nonumber \\
&& + Y(u_Z+t_Z)[u_ZC_Z^{211}(u)- u_hC_h^{211}(u)] \Big \}
+s\kappa (\mchil+\mchih)[t^2+u^2-2 m^2\mzd
\nonumber \\
&& +\kappa (t-u)]C_{hZ}^{121}(s) + 2s \Big \{ (\mchil+\mchih)
\Big [ s (t-u)(\mchil^2-\mchih^2)^2 -s u (u^2-m^2 \mzd)
\nonumber \\
&& +2 \mchil^2(t-u)Y+(\mchil^2-\mchih^2)s [Y-2 (u^2-m^2\mzd)]
\Big ]
\nonumber \\
&& +\kappa \{ (\mchil+\mchih)(\mchil^2-\mchih^2)^2 s
-2 u_Z u_h\mchil (\mchil^2-\mchih^2) +Y\mchil (\mchil^2+\mchih^2)
\nonumber \\
&& +\mchih s u (u-2 \mchih^2)-\mchil (u +2 \mchil\mchih)u_Zu_h \}
\Big \} D_{hZ}^{1211}(s,u)
\nonumber \\
&& + \frac{(\mchil+\mchih)}{4}\Big \{ 2 s (t-u)
[s (\mchil^2-\mchih^2)^2+ (\mchil^2+\mchih^2)Y]
\nonumber \\
&& +\kappa [2 s (\mchil-\mchih)^2+Y][s (\mchil+\mchih)^2+Y]
\Big \} [D_{hZ}^{1221}(t,u)+D_{hZ}^{2112}(t,u)]
\nonumber \\
&& -\frac{(\mchil-\mchih)Y}{4}(t_Z+u_Z)[s (\mchil+\mchih)^2+Y]
[D_{hZ}^{1221}(t,u)- D_{hZ}^{2112}(t,u)]
\nonumber \\
&& - (t \leftrightarrow   u ~~, ~~ \kappa \rightarrow  -\kappa ) \Bigg ]
~~ +  ~~ (1 \leftrightarrow 2) \Bigg \}
~~ , \label{h0-Zachi12-box+--}
\\
&& A^{Z_a \tchi_1\tchi_2 -\rm box}_{++0}(H) = \frac{2}{\kappa \mz s}
\Bigg \{ 4 \mchil s^2 (t+u) C_0^{111}(s) +
(\mchil+\mchih)[(t+u)(m^2+\mzd)
\nonumber \\
&& -4 m^2\mzd]E_2^{12}(t,u)- 2 s \mchil \Big (
\mchil \mchih \kappa^2 +2 s m^2 \mzd +\mchil^2
[(t+u)(m^2+\mzd)-4 m^2\mzd]
\nonumber \\
&& -s \mchih^2(t+u)\Big )[D_{hZ}^{1211}(s,u)+ D_{hZ}^{1211}(s,t)]
-(\mchil+\mchih) \Big (
s(\mchil^2+\mchih^2)[(t+u)(m^2+\mzd)
\nonumber \\
&& -4m^2\mzd]- 2 \mchil\mchih s^2 (t+u) \Big ) D_{hZ}^{1221}(t,u)
~~+~~ (1\leftrightarrow 2) \Bigg \}~~ , \label{h0-Zachi12-box++0}
\\
&& A^{Z_a \tchi_1\tchi_2 -\rm box}_{+-0}(H) =
\frac{(\mchil+\mchih)}{\kappa \mz Y} \Bigg \{ \Big \{
(t-u)(t_Z+u_Z)\Big [ s [t+u +2 (\mchil^2-\mchih^2)]C_0^{111}(s)
\nonumber \\
&& -\kappa^2 C_{hZ}^{121}(s) \Big ]-
2(t_Z+u_Z)(u^2-m^2\mzd)E_1^{12}(s,u)+4\mzd
Y[u_hC_h^{211}(u)-u_ZC_Z^{211}(u) ]
\nonumber \\
&& -2(t_Z+u_Z)\Big [ 2 \mchil^2 (t-u)Y+s(t-u)(\mchil^2-\mchih^2)^2
\nonumber \\
&&+s(\mchil^2-\mchih^2)[Y-2(u^2-m^2\mzd)]\Big ] D_{hZ}^{1211}(s,u)
-\frac{(t-u)}{2}(t_Z+u_Z)[s(\mchil^2-\mchih^2)^2
\nonumber \\
&&+Y(\mchil^2+\mchih^2)]
[D_{hZ}^{1221}(t,u)+D_{hZ}^{2112}(t,u)]
\nonumber \\
&& +\frac{\mchil-\mchih}{\mchil+\mchih}\mzd Y[Y
+s (\mchil+\mchih)^2]
[D_{hZ}^{1221}(t,u)
 -D_{hZ}^{2112}(t,u)]
\nonumber \\
&&  ~~-~~ (t\leftrightarrow  u) \Big \}
~~+~~ (1  \leftrightarrow 2) \Bigg \}
~~ . \label{h0-Zachi12-box+-0}
\eqa

\vspace{1cm}
{\bf \underline{Mixed chargino box involving vector Z coupling.}}
The generic form of these box diagrams is
shown in Fig.\ref{ggZh0-diag}f,g, which  are analogous
to those in   (d, e);  but
involve  the vector  $Z\tchi_1\tchi_2$-couplings in
(\ref{Z-chi12-coupling}),
combined with the $g_{p12}^{h^0}$, $g_{p12}^{H^0}$  Higgs
ones of (\ref{H-chi12-coupling}).
The contribution of these diagrams
may be obtained from those  of Fig.\ref{ggZh0-diag}d,e by simply changing
the sign of one chargino mass.
More explicitly, if we write
\bq
F^{Z_v \tchi_1\tchi_2 -\rm box}_{\lambda_1\lambda_2\lambda_Z}
(\gamma \gamma \to Z h^0(H^0) )=
r_{Z_v \chi_1\chi_2}^{h^0(H^0)} \cdot
A^{Z_v \tchi_1\tchi_2 -\rm box}_{\lambda_1\lambda_2\lambda_Z}(H) ~~ ,
\label{Zvchi12-box-amplitude-h0}
\eq
where the relevant  couplings defined in
(\ref{Z-chi12-coupling}, \ref{H-chi12-coupling}),
 are absorbed in the coefficients
\bqa
r_{Z_v \chi_1\chi_2}^{h^0}&=&
- \frac{e^3}{(4\pi)^2} g_{v12}^Z g_{p12}^{h^0} ~~ ,
\nonumber \\
r_{Z_v \chi_1\chi_2}^{H^0}&=&
- \frac{e^3}{(4\pi)^2} g_{v12}^Z g_{p12}^{H^0} ~~ ,
\label{r-Zv-chi12-box-H}
\eqa
and the amplitudes
$A^{Z_v \tchi_1\tchi_2 -\rm box}_{\lambda_1\lambda_2\lambda_Z}$
of (\ref{Zvchi12-box-amplitude-h0}) are determined by
(\ref{h0-Zachi12-box+++}-\ref{h0-Zachi12-box+-0}) through
\bq
A^{Z_v \tchi_1\tchi_2 -\rm box}_{\lambda_1\lambda_2\lambda_Z}
(H, \mchil, \mchih) =
A^{Z_a \tchi_1\tchi_2 -\rm box}_{\lambda_1\lambda_2\lambda_Z}
(H, \mchil, - \mchih)
=- A^{Z_a \tchi_1\tchi_2 -\rm box}_{\lambda_1\lambda_2\lambda_Z}
(H, - \mchil, \mchih) ~. \label{Zvchi12-box-amplitude1-h0}
\eq
Notice that the  constraint on
$A^{Z_a \tchi_1\tchi_2 -\rm box}_{\lambda_1\lambda_2\lambda_Z}
(H, \mchil, \mchih)$ implied by (\ref{Zvchi12-box-amplitude1-h0}),
is satisfied by the expressions in
(\ref{h0-Zachi12-box+++}-\ref{h0-Zachi12-box+-0}).\\

Concerning the SM case $\gamma \gamma \to Z H_{SM}$,
we note that it can be obtained from   (\ref{Z-pole-tbtau-h0},
\ref{f-box-amplitude-h0}), by replacing $h^0 \to H_{SM}$ and using
 $\alpha=\beta-\pi/2$.

\vspace{2cm}

\newpage

\renewcommand{\theequation}{C.\arabic{equation}}
\renewcommand{\thesection}{C.\arabic{section}}
\setcounter{equation}{0}
\setcounter{section}{0}

{\large \bf Appendix C:  The $\gamma \gamma \to Z A^0$
helicity amplitudes.}

The  helicity amplitudes for
$\gamma \gamma \to Z A^0 $
\bq
\gamma (k_1,\lambda_1) \gamma (k_2,\lambda_2) \to
Z (q_1,\lambda_Z) ~A^0 (q_2) \ \ ,
\label{ggZA0-process}
\eq
denoted  again  as
$F_{\lambda_1 \lambda_2 \lambda_Z}(\kappa ,t,u)$,
should satisfy the constraints
\bqa
F_{\lambda_1, \lambda_2, \lambda_Z}(\kappa ,t,u) & = &
F_{-\lambda_1, -\lambda_2, -\lambda_Z}(\kappa
,t,u)(-1)^{\lambda_Z} ~ ,\label{CP-A0-con} \\
F_{\lambda_1 \lambda_2 \lambda_Z}(\kappa ,t,u) &= &
F_{\lambda_2 \lambda_1 \lambda_Z}(\kappa ,u,t)(-1)^{\lambda_Z} ~,
\label{Bose-A0-con} \\
F_{\lambda_1, \lambda_2, \lambda_Z}(\kappa ,t,u) &= & -
F_{\lambda_1, \lambda_2, -\lambda_Z}(-\kappa
,t,u)(-1)^{\lambda_Z} ~ ,\label{Zpol-A0-con}
\eqa
imposed respectively by CP-invariance at the 1-loop level,
Bose statistics and
the structure of the Z-polarization vector. Thus,
for the $A^0$ production case also, there are
only four "basic" helicity amplitudes which are  taken as
\bq
F_{+++}~~~,~~~ F_{+--}~~~,~~~ F_{++0}~~~,~~~ F_{+-0} ~~~.
\label{basic-A0-amplitudes}
\eq
Because of (\ref{CP-A0-con}-\ref{Zpol-A0-con}), the $A^0$
production amplitudes  still obey
(\ref{ampl-con1}, \ref{ampl-con2}), but (\ref{ampl-con3})
is modified to
\bqa
&& F_{+-0}(\kappa, t, u)=- F_{+-0}(-\kappa, t, u)=
 F_{-+0}(\kappa, t, u)
\nonumber \\
&& = F_{+-0}(\kappa, u, t)= F_{-+0}(\kappa, u, t) ~~ .
\label{ampl-con4}
\eqa\par

\begin{figure}[p]
\vspace*{-2cm}
\[
\epsfig{file=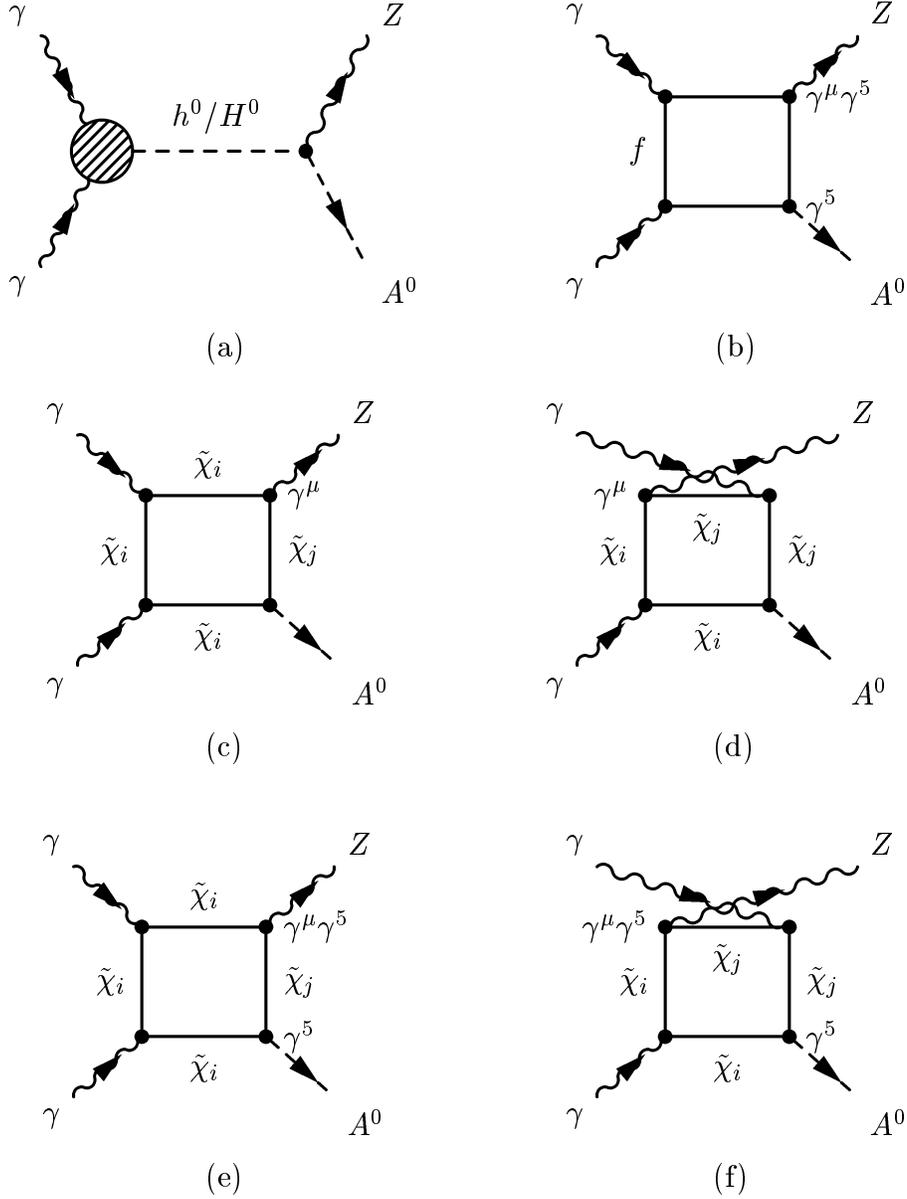,height=16cm,width=12cm}
\]
\caption[1]{Generic diagrams describing the various
contributions to the $\gamma \gamma \to Z A^0$ in
SUSY models.  Solid lines correspond to fermions, broken ones to scalars,
while wavy lines correspond to gauge bosons. The diagrams
in (c-f) for $j \neq i$ describe the mixed chargino boxes. }
\label{ggZA0-diag}
\end{figure}

\begin{figure}[p]
\vspace*{0cm}
\[
\hspace{-0.5cm}\epsfig{file=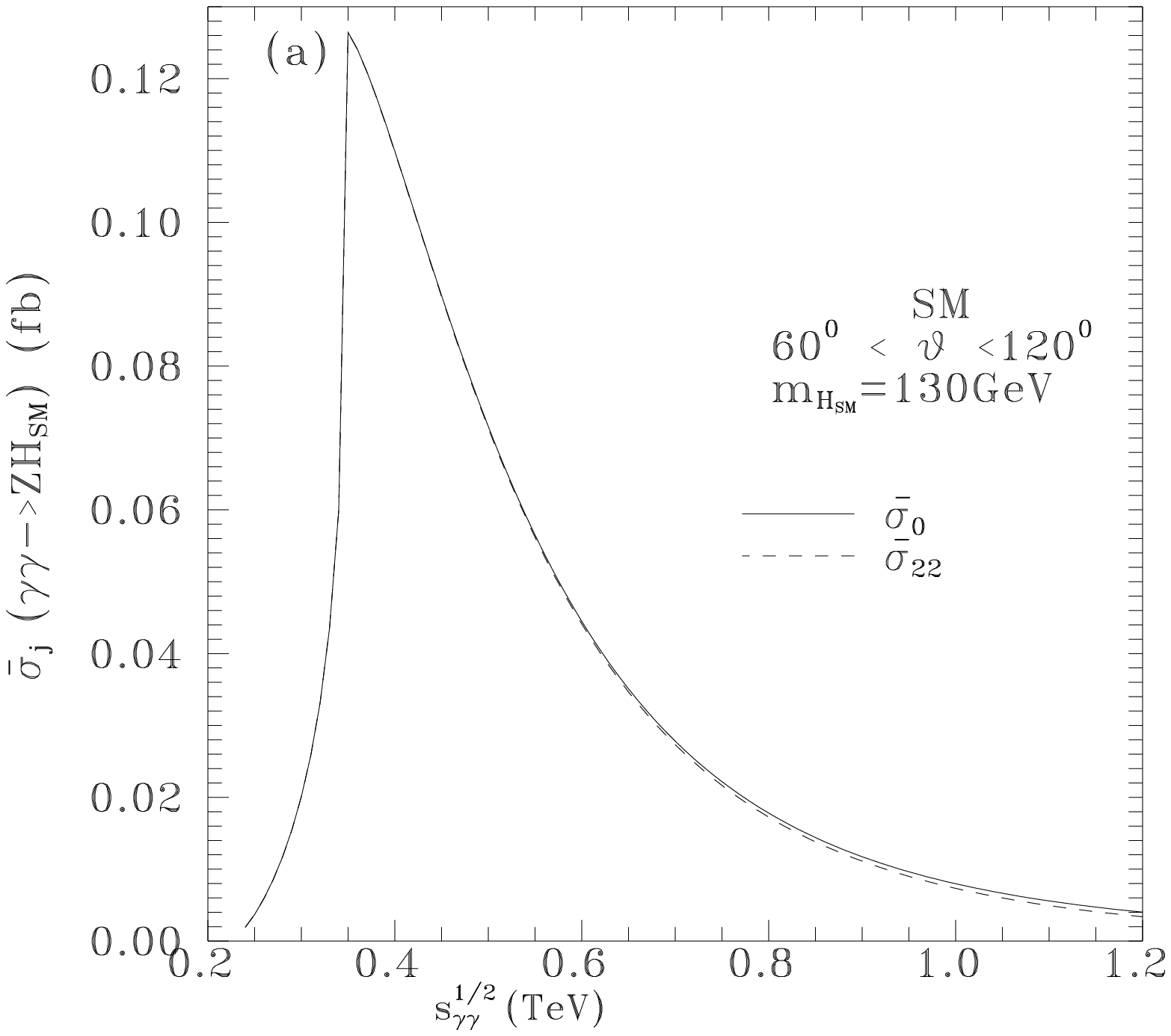,height=7cm}\hspace{0.5cm}
\epsfig{file=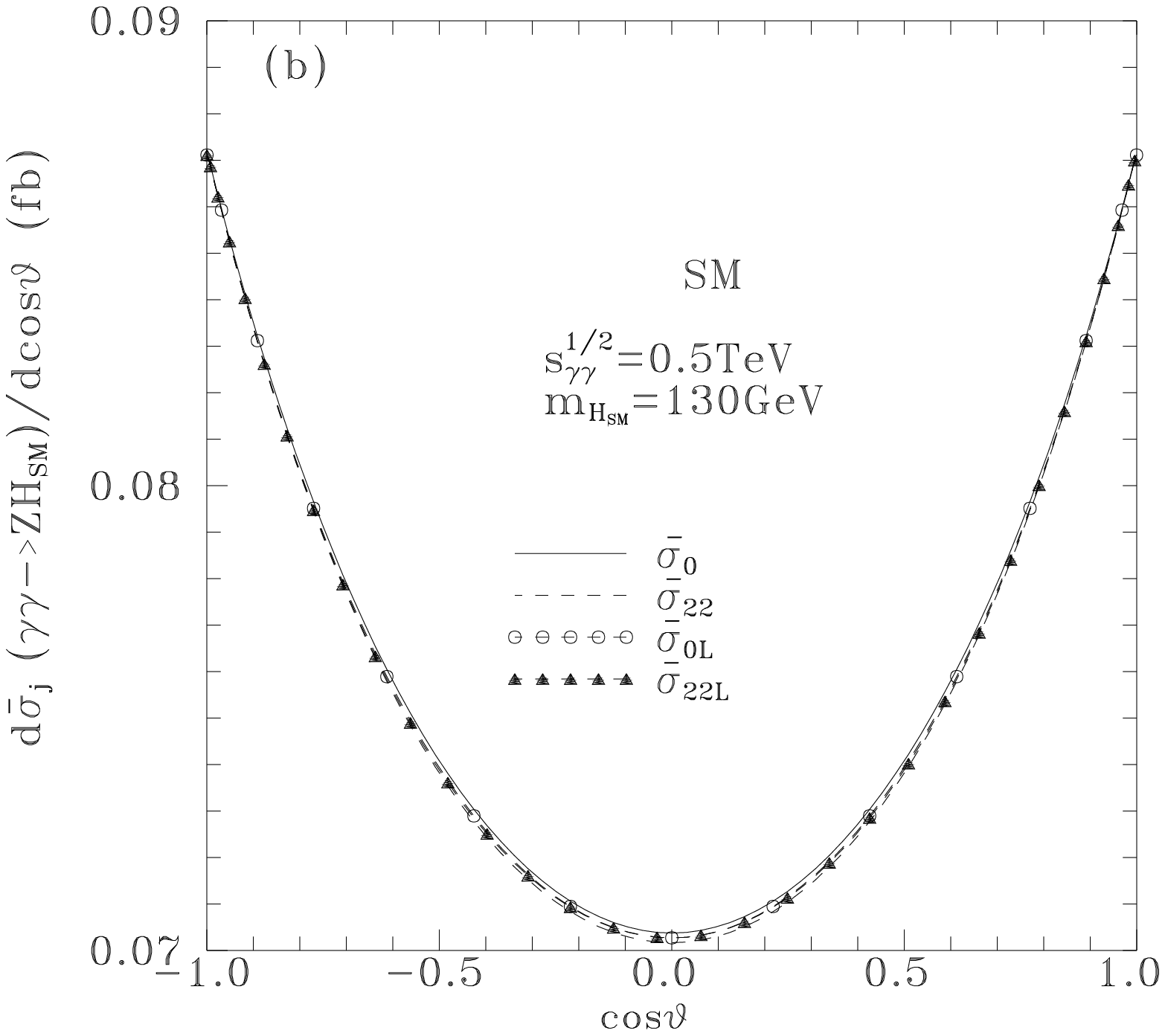,height=7cm}
\]
\caption[1]{$\gamma \gamma \to Z H$
cross sections in SM. The cross sections $\bar{\sigma}_{0L}$ and
$\bar{\sigma}_{22L}$ refer to the production of longitudinal $Z$
bosons.}
\label{ggZH-SM-fig}
\end{figure}

\begin{figure}[p]
\vspace*{-3cm}
\[
\hspace{-0.5cm}\epsfig{file=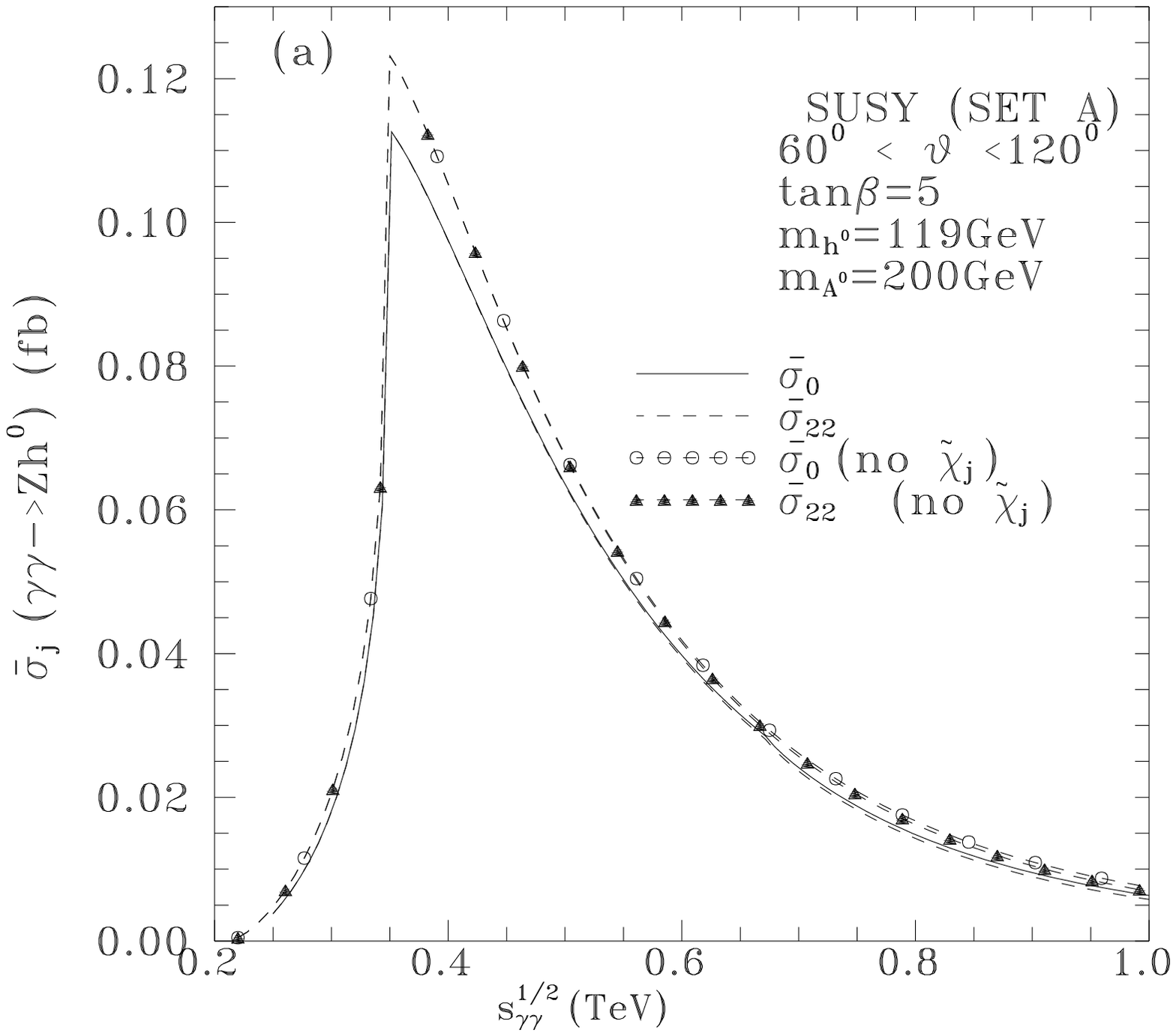,height=7cm}\hspace{0.5cm}
\epsfig{file=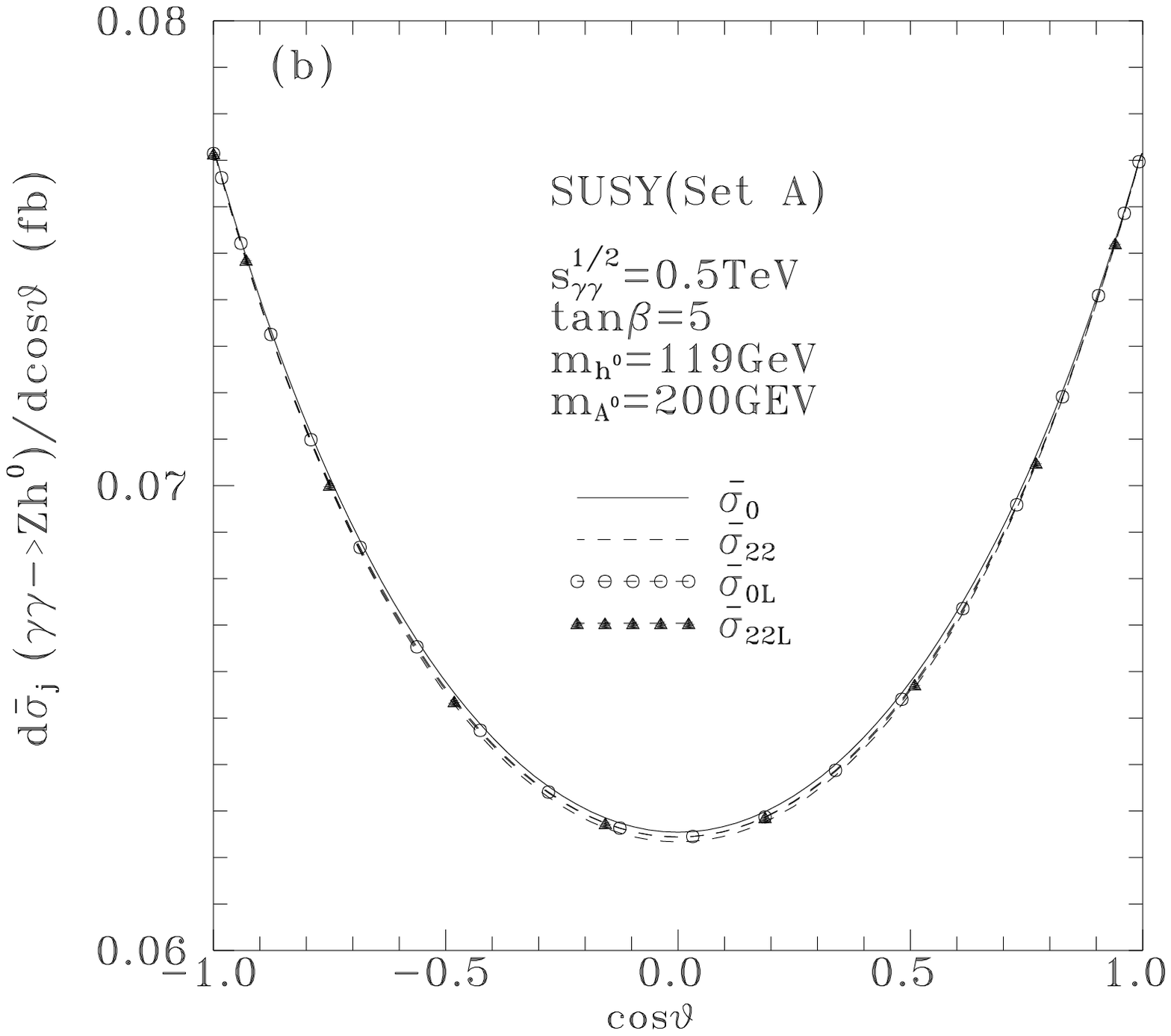,height=7cm}
\]

\vspace*{0.5cm}
\[
\hspace{-0.5cm}\epsfig{file=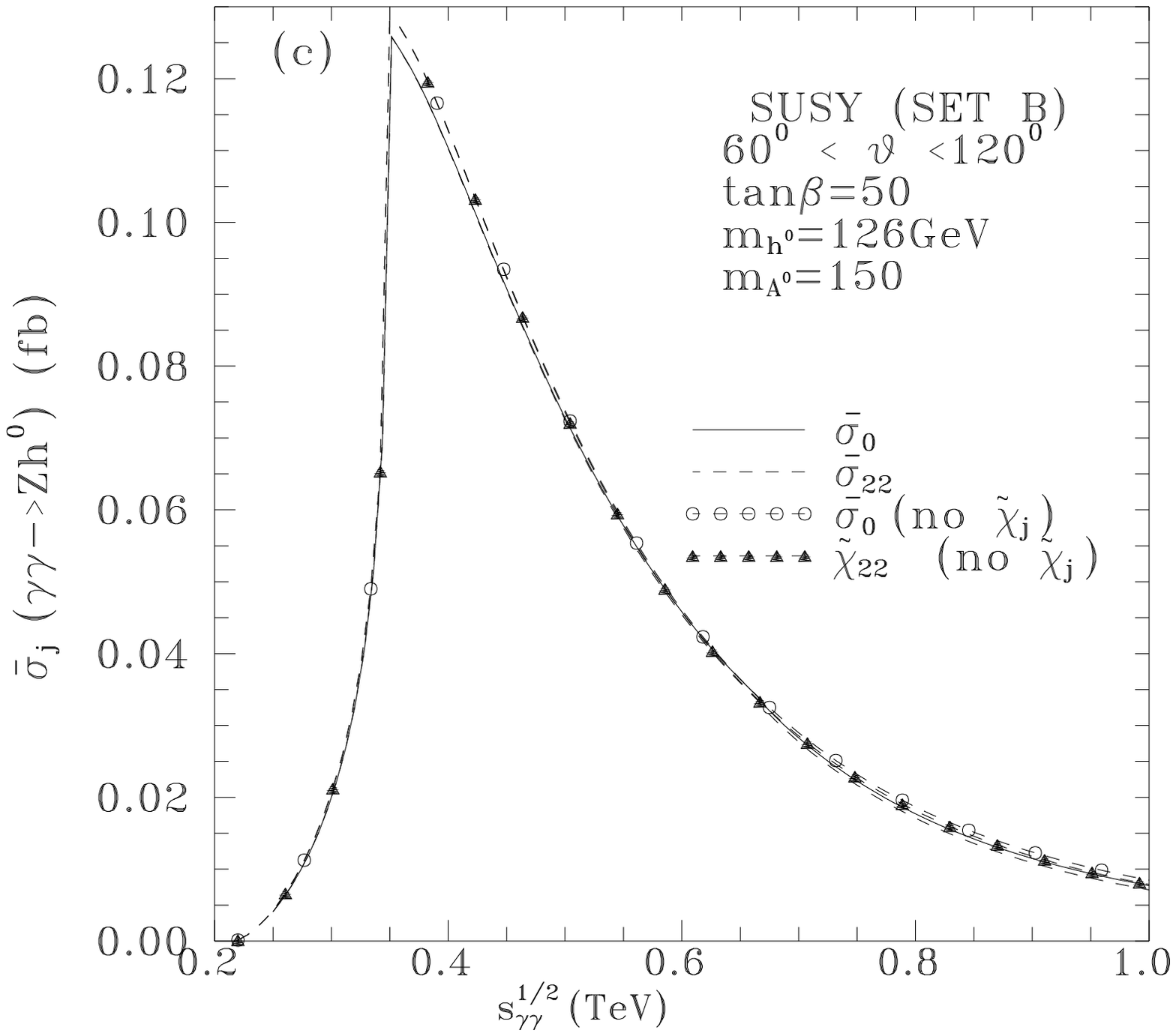,height=7cm}\hspace{0.5cm}
\epsfig{file=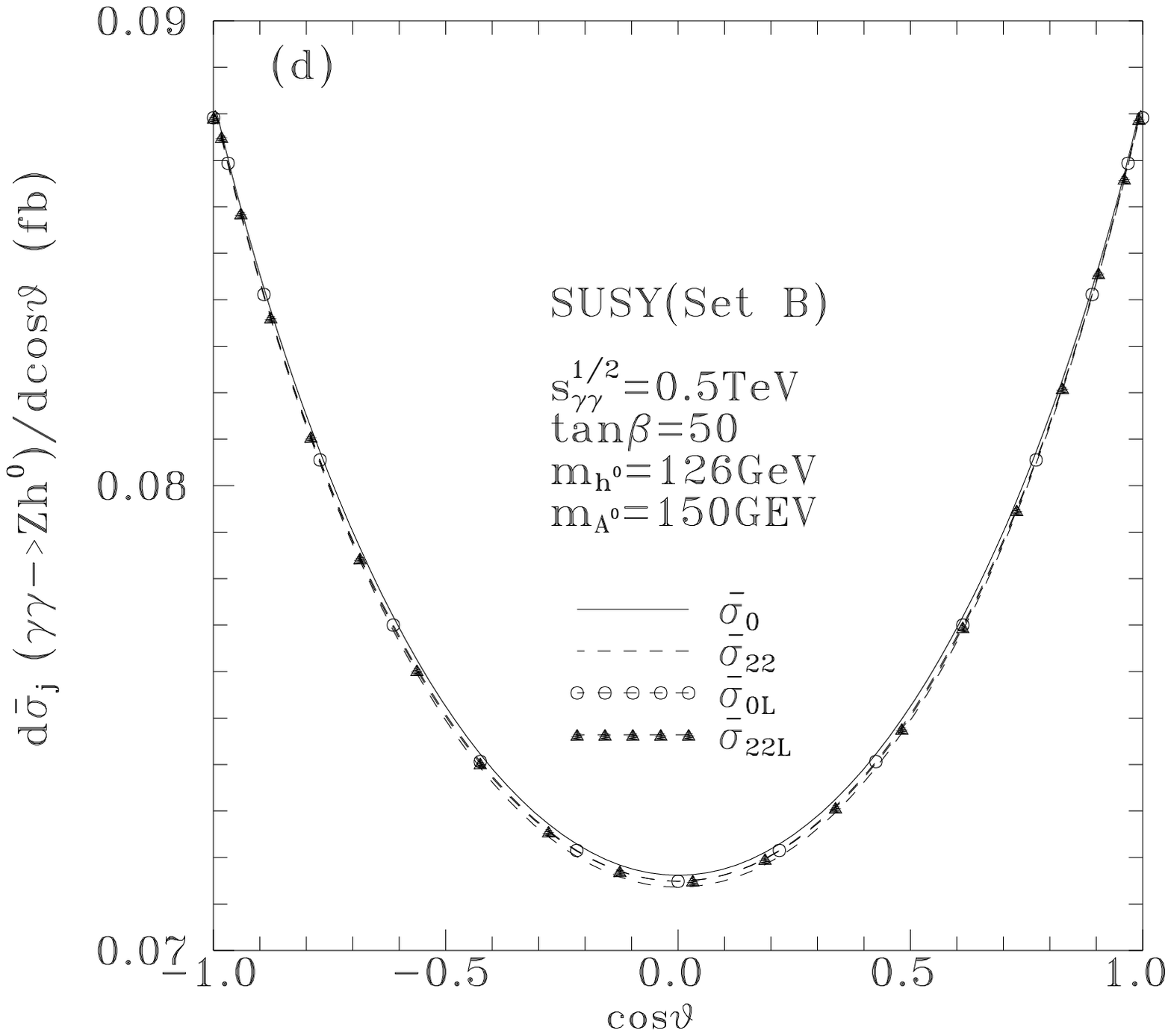,height=7cm}
\]
\vspace*{0.5cm}
\caption[1]{$\gamma \gamma \to Z h^0$
cross sections in SUSY. The complete list of the parameters used
in sets A and B appear in Table 1. The label (no $\tilde{\chi}_j$)
means that the chargino contribution has been suppressed in the
computation of the cross section.}
\label{ggZh0-SUSY-fig}
\end{figure}

\begin{figure}[p]
\vspace*{-3cm}
\[
\hspace{-0.5cm}\epsfig{file=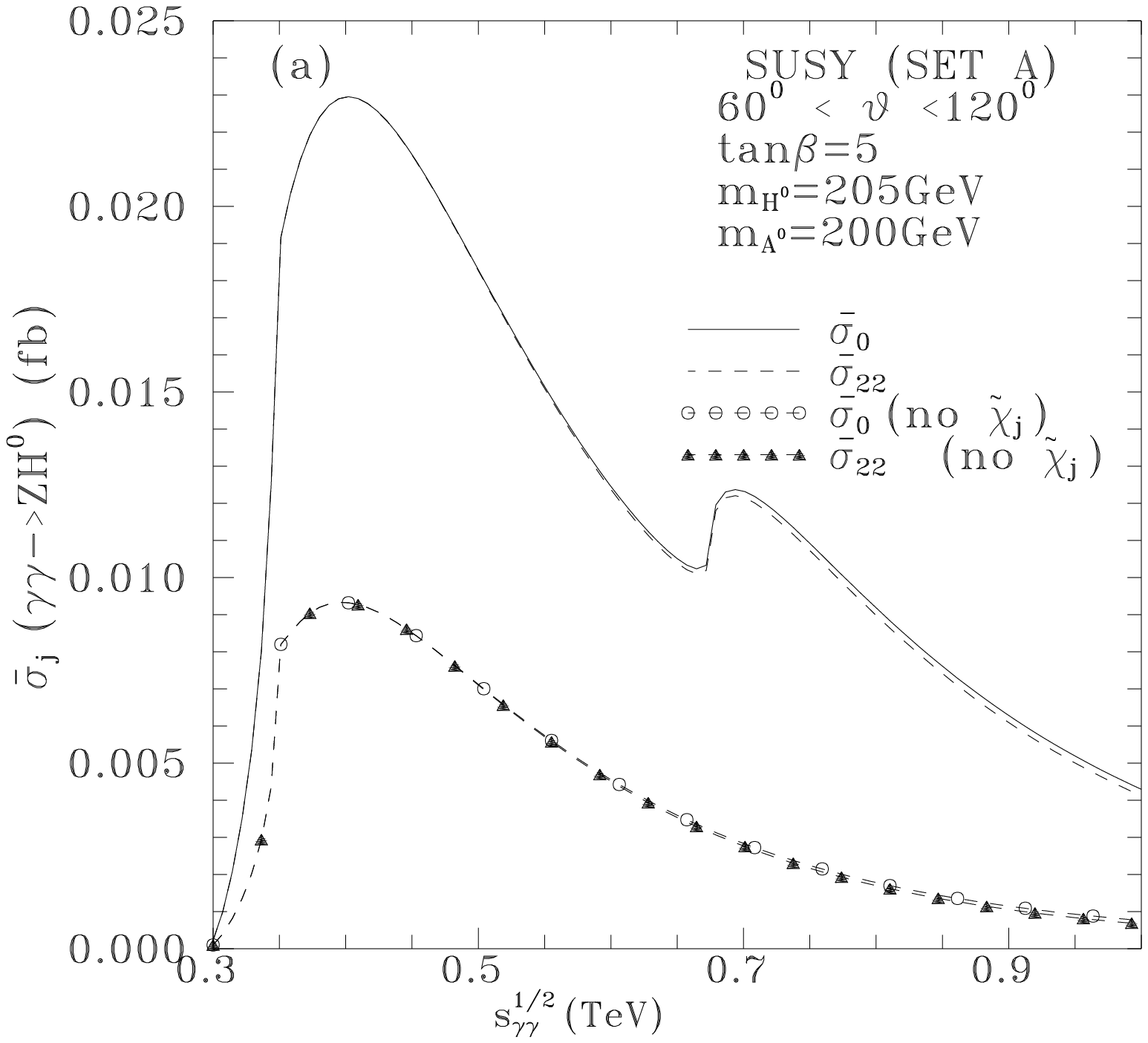,height=7cm}\hspace{0.5cm}
\epsfig{file=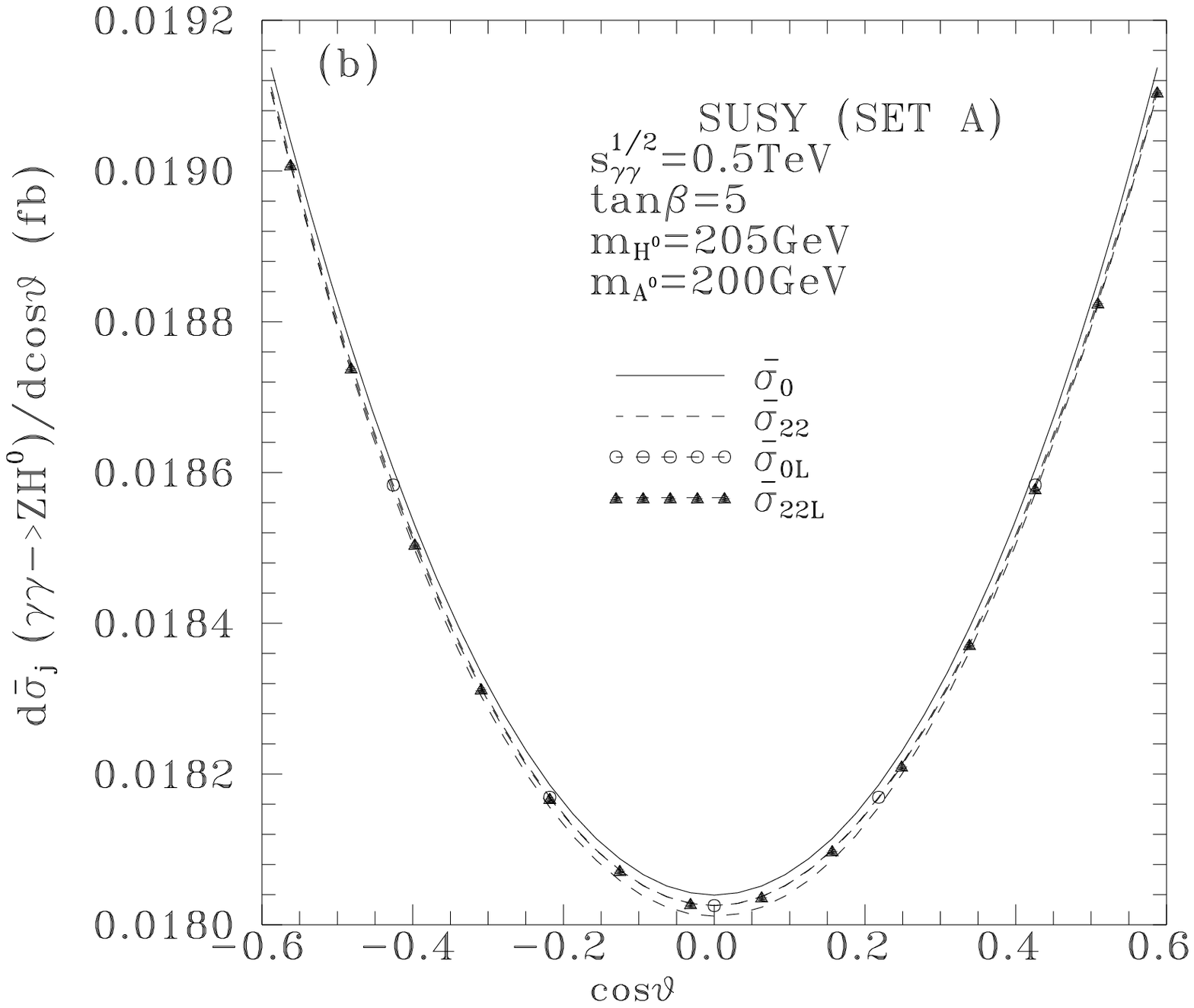,height=7cm}
\]

\vspace*{0.5cm}
\[
\hspace{-0.5cm}\epsfig{file=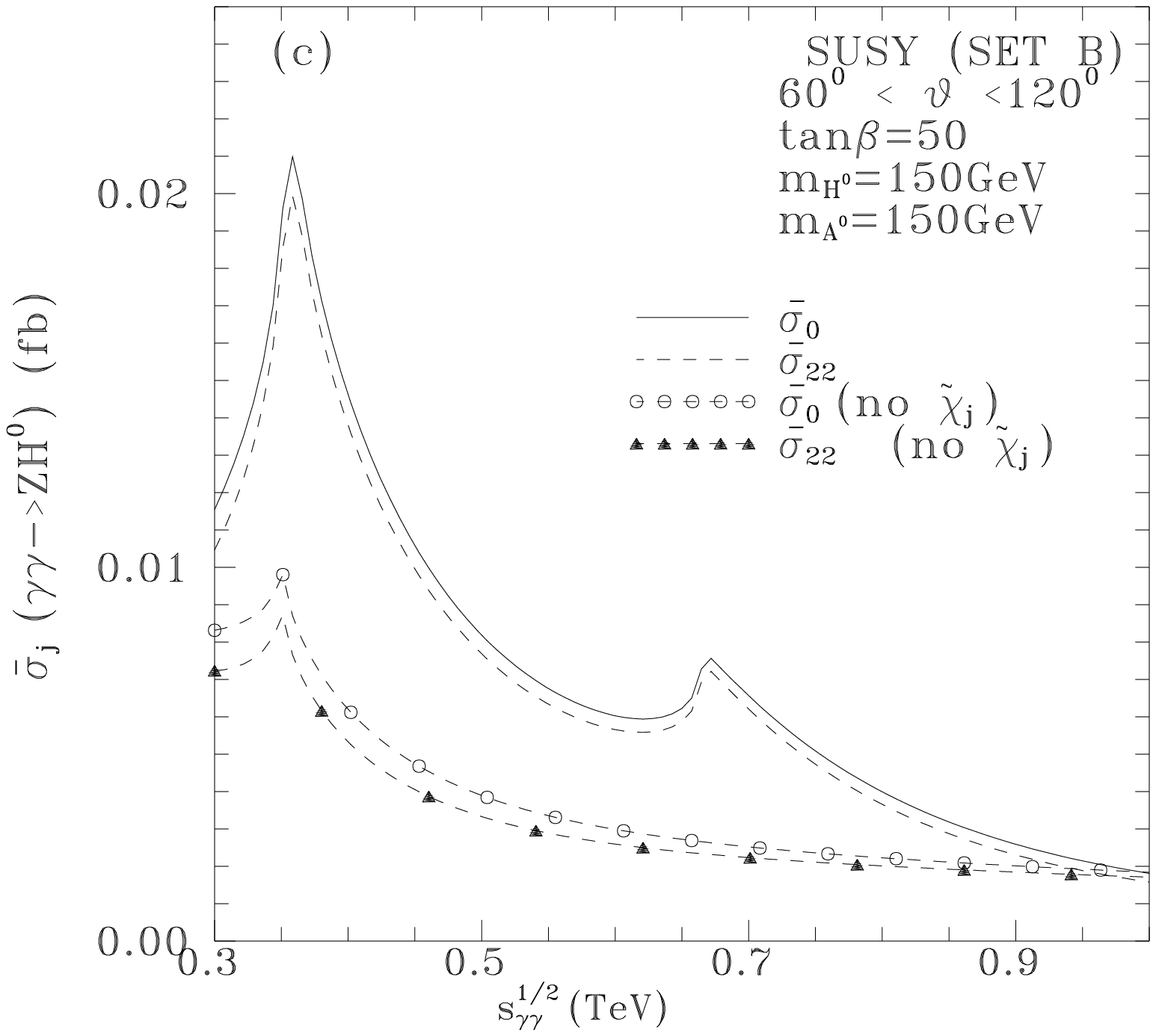,height=7cm}\hspace{0.5cm}
\epsfig{file=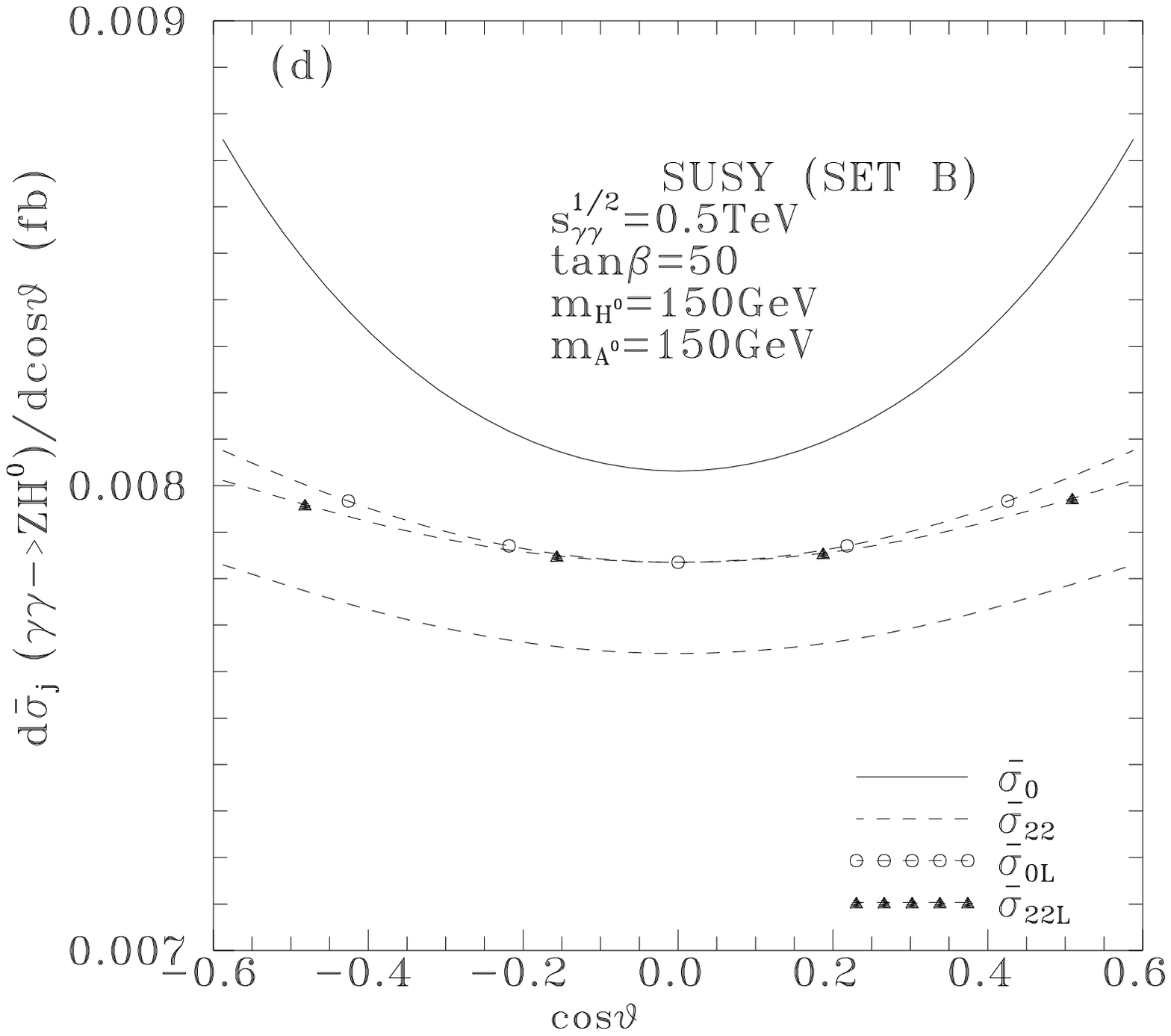,height=7cm}
\]
\vspace*{0.5cm}
\caption[1]{$\gamma \gamma \to Z H^0$
cross sections in SUSY. The complete list of the parameters used
appear   in Table 1. Same captions as in Fig.3,4.}
\label{ggZH0-SUSY-fig}
\end{figure}

\begin{figure}[p]
\vspace*{-3cm}
\[
\hspace{-0.5cm}\epsfig{file=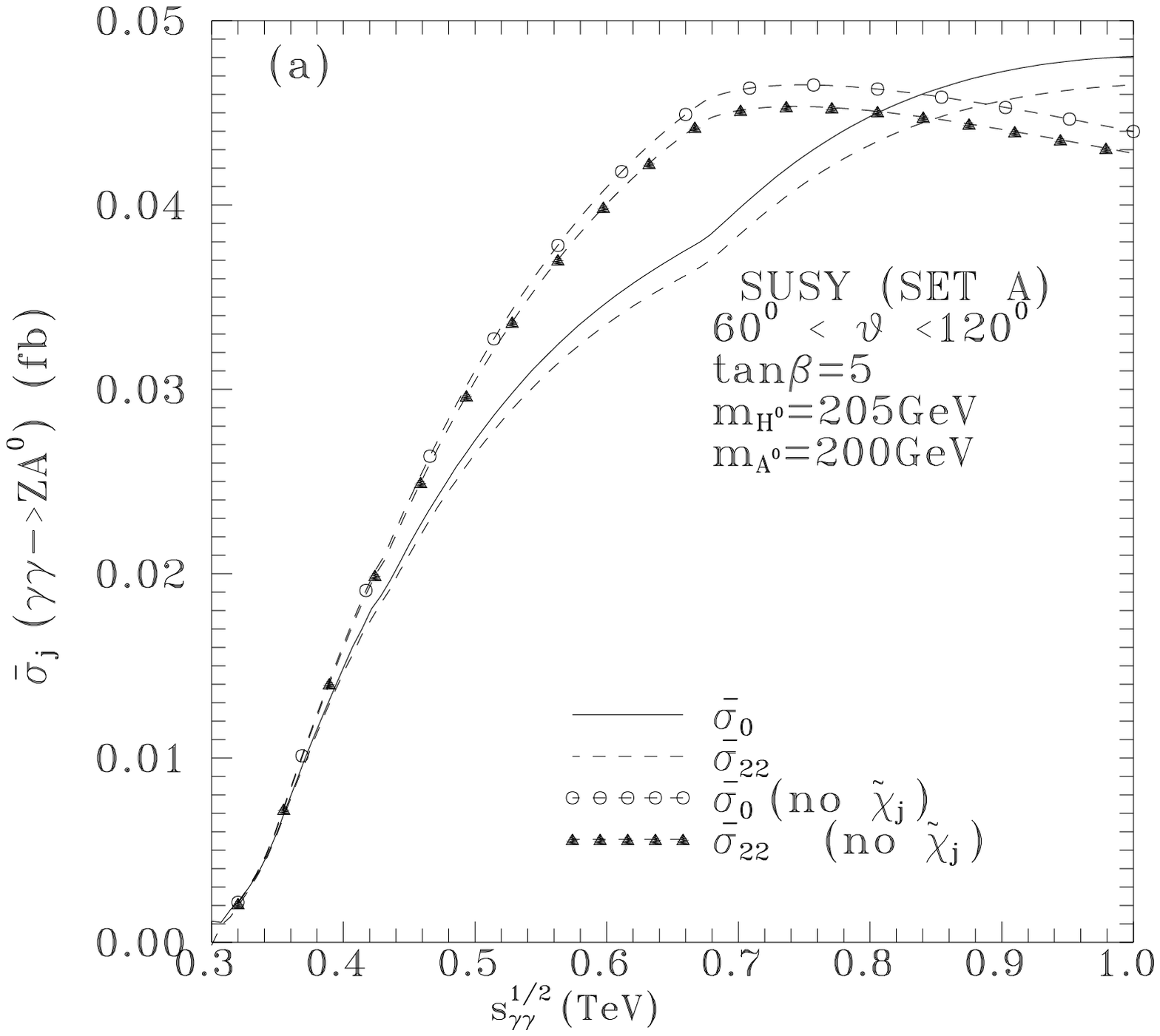,height=7cm}\hspace{0.5cm}
\epsfig{file=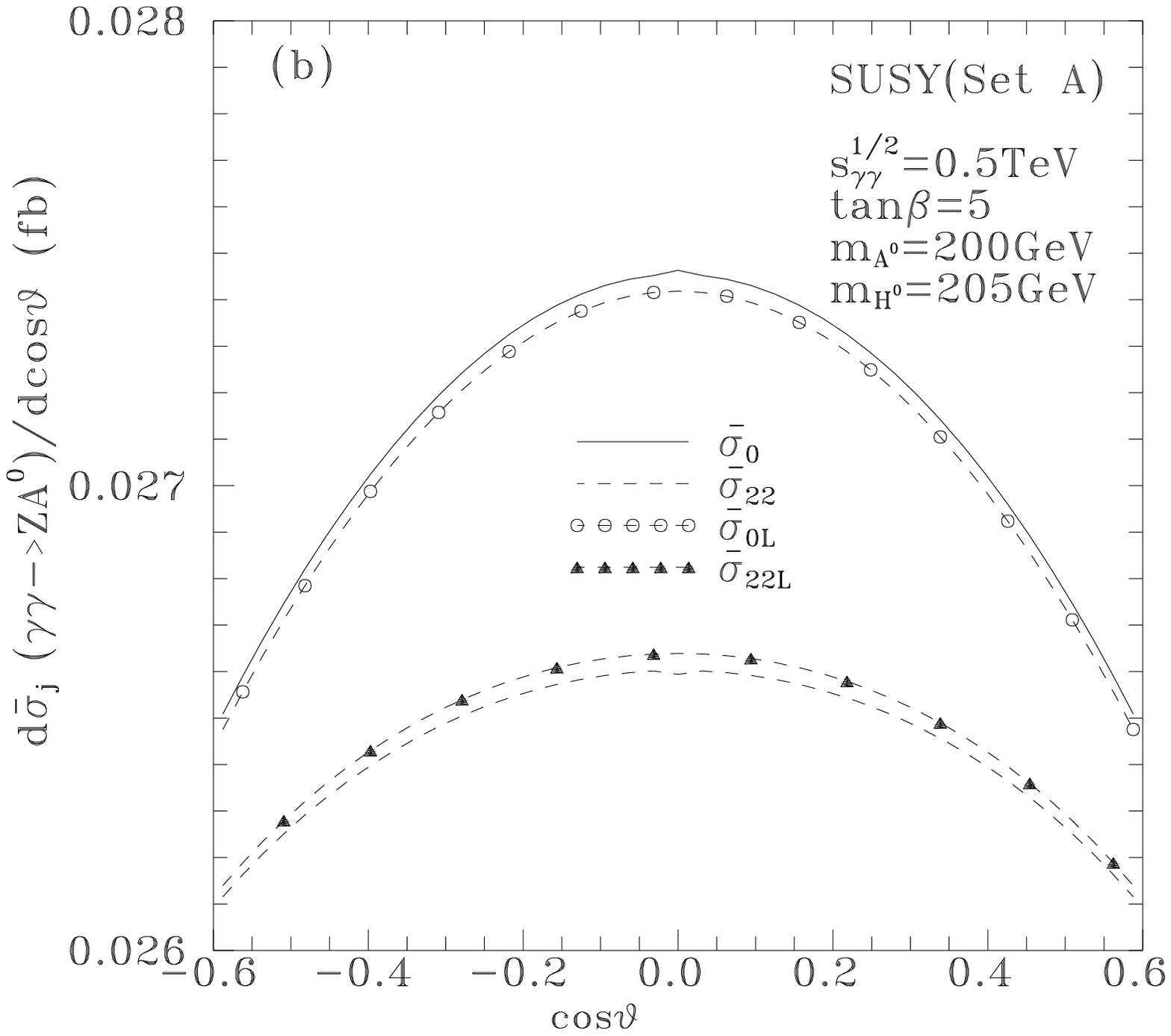,height=7cm}
\]

\vspace*{0.5cm}
\[
\hspace{-0.5cm}\epsfig{file=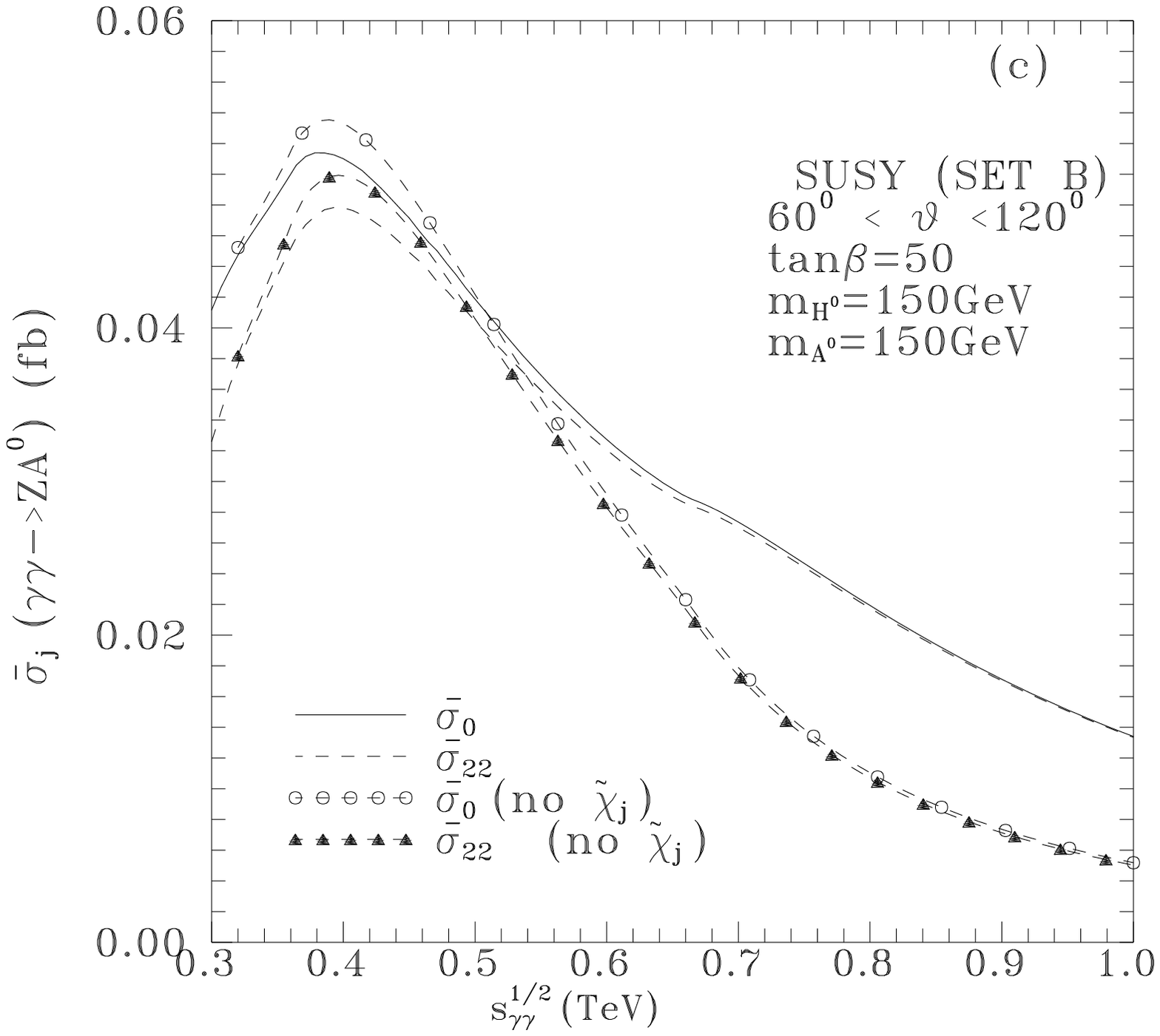,height=7cm}\hspace{0.5cm}
\epsfig{file=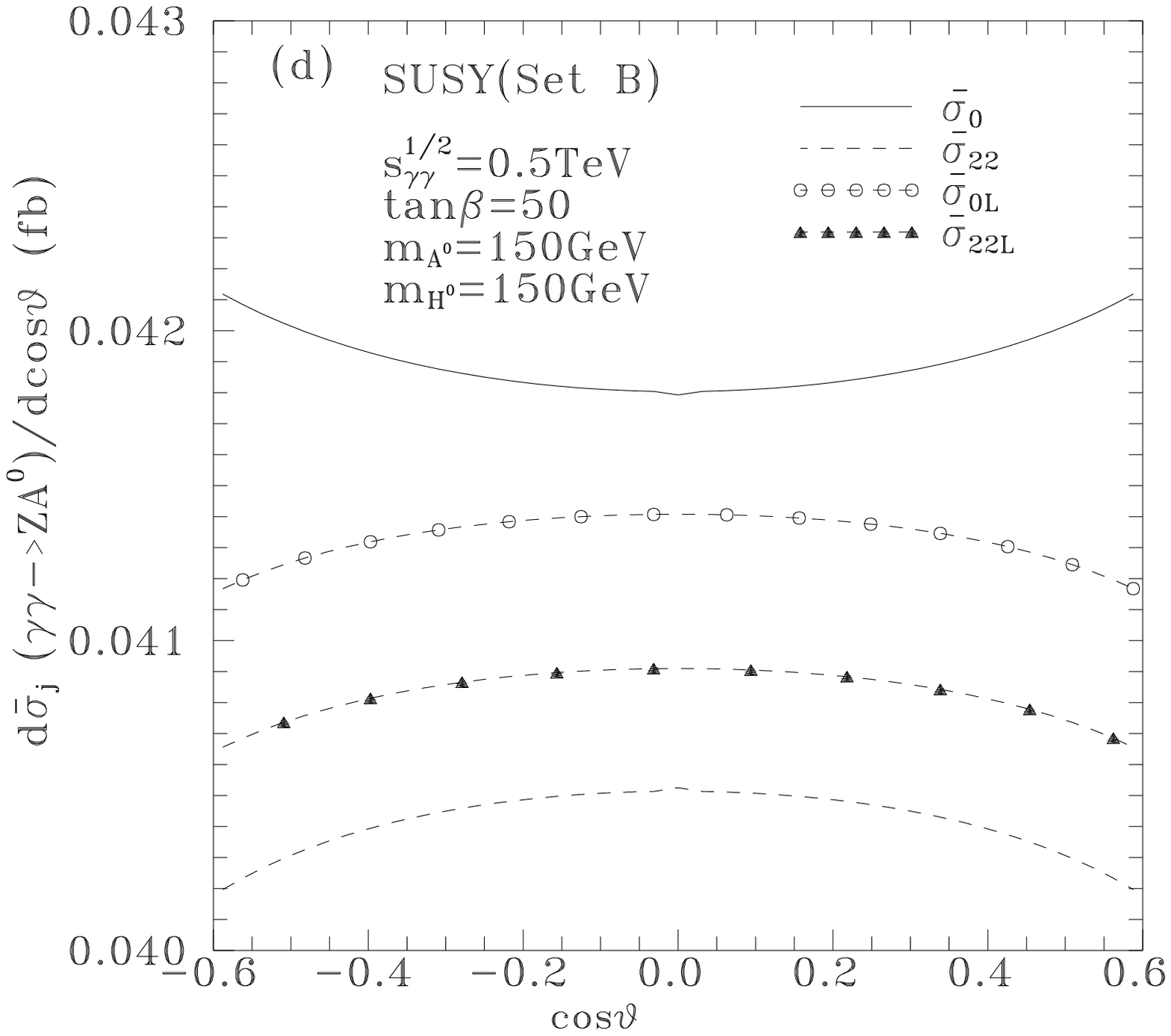,height=7cm}
\]
\vspace*{0.5cm}
\caption[1]{$\gamma \gamma \to Z A^0$
cross sections in SUSY. The complete list of the parameters used
appear   in Table 1. Same captions as in Fig.3,4.}
\label{ggZA0-SUSY-fig}
\end{figure}

The relevant diagrams are shown in Fig.\ref{ggZA0-diag}.
Below we discuss their respective contributions.\par

\vspace{1cm}
{\bf \underline{The $h^0~,~H^0$ pole contribution.}}
This is described by the diagram in Fig.\ref{ggZA0-diag}a,
where  the blob denotes  loops from fermions, W-bosons and
scalars.\par

As in (\ref{A0-pole-f-h0}), the only non-vanishing
contribution from this diagram is for the $F_{++0}$ amplitude.
The fermion loop contributions to it is
\bqa
F^{(h^0,H^0)f-\rm pole}_{++0}(\gamma \gamma \to Z A^0)
&= & - \frac{i \alpha g \kappa m_f Q_f^2 N_f^c}{2 \cw \mz\pi}
[(s-4m_f^2)C_0^f(s)-2]\Big [
\frac{g_{h^0ff}\cos(\beta-\alpha)}{s-m_{h^0}^2}
\nonumber \\
& - &  \frac{g_{H^0ff}\sin(\beta-\alpha)}
{s-m_{H^0}^2 +i m_{H^0} \Gamma_{H^0}} \Big ] ~~,
\label{H-pole-f-A0}
\eqa
where the values of the $g_{h^0ff},~ g_{H^0ff}$ couplings for
the 3rd family fermions   $(t,~ b, ~ \tau)$ are given in
(\ref{H-f-coupling}). The same
relation (\ref{H-pole-f-A0}) describes also the chargino loop
contribution to the $\gamma \gamma h^0 (H^0)$ vertex, provided
$(g_{h^0ff},~ g_{H^0ff})\rightarrow (g_j^{h^0},~ g_j^{H^0})$, with
the later couplings given in (\ref{H-chi1-coupling},
\ref{H-chi2-coupling}).\par

For the $W$ (plus Goldstone and ghost) contribution to the blob in
Fig.\ref{ggZA0-diag}a, we have
\bqa
&&F^{(h^0,H^0)W-\rm pole}_{++0}(\gamma \gamma \to Z A^0)=
\nonumber \\
&& \frac{i \alpha^2 \kappa}{2\swd} \Bigg \{
\left ( \frac{1}{s-m_{h^0}^2}- \frac{1}
{s-m_{H^0}^2}
\right ) \left [ 3-(4s-6\mwd)C_0^W(s)\right ]
\sin 2(\beta-\alpha)
\nonumber \\
&&  - \frac{\cos(2\beta)}{\cwd}[1+2\mwd C_0^W(s)]
\Big [ \frac{\cos(\beta-\alpha)\sin(\beta+\alpha)}{s-m_{h^0}^2}
 +\frac{\sin(\beta-\alpha)\cos(\beta+\alpha)}
{s-m_{H^0}^2} \Big ]\Bigg \}.
\label{H-pole-W-A0}
\eqa\par

Finally, the scalar contribution in the $\gamma \gamma h^0 (H^0)$
vertices give
\bq
F^{(h^0,H^0)\rm scalar- pole}_{++0}(\gamma \gamma \to Z A^0)=
\frac{i \alpha^2 \kappa}{4 \swd \mwd} {\cal H}^{\rm scalar}(s)
~, \label{H-pole-scalar-A0}
\eq
where for the charged Higgs loop we have
\bqa &&
{\cal H}^{H^+}(s)= 4 \mwd \left [1+2 m_{H^+}^2 C_0^{H^+}(s)\right ]\Bigg
\{ \frac{\cos(\beta-\alpha)}{s-m_{h^0}^2}\Big [\sin(\beta-\alpha)
+\frac{\cos(2\beta)\sin(\alpha+\beta)}{2\cwd} \Big ] \nonumber \\
&& - \frac{\sin(\beta-\alpha)}{s-m_{H^0}^2}\Big
[\cos(\beta-\alpha) -\frac{\cos(2\beta)\cos(\alpha+\beta)}{2\cwd}
\Big ]\Bigg \} ~ , \label{H-pole-cHiggs-A0}
\eqa
for the lighter stop $\stop_1$ loop
\bqa
&& {\cal H}^{\stop_1}(s)= 4 \left [1+2
m_{\stop_1}^2 C_0^{\stop_1}(s)\right ]\Bigg \{
\frac{\cos(\beta-\alpha)}{s-m_{h^0}^2} \Big [
-\frac{\mwd}{\cwd}\sin(\alpha+\beta)\Big [\frac{2\swd}{3}
+\Big (\frac{1}{2}-\frac{4\swd}{3}\Big ) \cos^2\theta_t \Big ]
\nonumber \\
&& +\frac{\mt^2 \calpha}{\sbeta}
 +\frac{\mt (A_t\calpha +\mu \salpha)}{2\sbeta}
\sin(2\theta_t) \Sn(A_t-\mu \cot\beta) \Big ]
\nonumber \\
&& -\frac{\sin(\beta-\alpha)}{s-m_{H^0}^2} \Big [
\frac{\mwd}{\cwd}\cos(\alpha+\beta)\Big [\frac{2\swd}{3}
+\Big (\frac{1}{2}-\frac{4\swd}{3}\Big ) \cos^2\theta_t \Big ]
 +\frac{\mt^2 \salpha}{\sbeta}
\nonumber \\
&&  +\frac{\mt (A_t\salpha -\mu \calpha)}{2\sbeta}
\sin(2\theta_t) \Sn(A_t-\mu \cot\beta) \Big ] \Bigg \} ~~,
\label{H-pole-stop1-A0}
\eqa
while for the $\stop_2$-loop contribution we get
\bqa
&& {\cal H}^{\stop_2}(s)= 4 \left [1+2
m_{\stop_2}^2 C_0^{\stop_2}(s)\right ]\Bigg \{
\frac{\cos(\beta-\alpha)}{s-m_{h^0}^2} \Big [
-\frac{\mwd}{\cwd}\sin(\alpha+\beta)\Big [\frac{2\swd}{3}
+\Big (\frac{1}{2}-\frac{4\swd}{3}\Big ) \sin^2\theta_t \Big ]
\nonumber \\
&& +\frac{\mt^2 \calpha}{\sbeta}
 -\frac{\mt (A_t\calpha +\mu \salpha)}{2\sbeta}
\sin(2\theta_t) \Sn(A_t-\mu \cot\beta) \Big ]
\nonumber \\
&& -\frac{\sin(\beta-\alpha)}{s-m_{H^0}^2} \Big [
\frac{\mwd}{\cwd}\cos(\alpha+\beta)\Big [\frac{2\swd}{3}
+\Big (\frac{1}{2}-\frac{4\swd}{3}\Big ) \sin^2\theta_t \Big ]
 +\frac{\mt^2 \salpha}{\sbeta}
\nonumber \\
&&  -\frac{\mt (A_t\salpha -\mu \calpha)}{2\sbeta}
\sin(2\theta_t) \Sn(A_t-\mu \cot\beta) \Big ] \Bigg \} ~~,
\label{H-pole-stop2-A0}
\eqa
where the various stop-parameters are defined as in
\cite{ggZZ-2nd}.

\vspace{1cm}
{\bf \underline{Singe fermion box contribution.}}
It is given by the diagram in
Fig.\ref{ggZA0-diag}b which is closely related to the diagram in
 Fig.\ref{ggZh0-diag}c for  the $h^0,~H^0$ production case.
In both cases,   the Z-coupling to fermions is axial, while
the main difference stems  from the $\gamma_5$ in the Higgs vertex
of  Fig.\ref{ggZA0-diag}b. In analogy to
(\ref{f-box-amplitude-h0}), the contribution of the diagram
in Fig.\ref{ggZA0-diag}b may be written  as
\bq
F^{f-\rm box}_{\lambda_1\lambda_2\lambda_Z}(\gamma \gamma \to
Z A^0 )= r_f^{A^0 }\cdot ~
A^{f-\rm box}_{\lambda_1\lambda_2\lambda_Z}(A^0) ~~ ,
\label{f-box-amplitude-A0}
\eq
with
\bqa
r_f^{A^0}&=& i
\frac{e^3}{(4\pi)^2}N_f^c Q_f^2 g_{af}^Z \tilde g_{A^0ff} ~~,
\nonumber \\
r_{\tchi_j}^{A^0}&=& i \frac{e^3}{(4\pi)^2}g_{aj}^Z
 \tilde g_j^{A^0} ~~ \label{r-f-box-A0}
\eqa
for  $(t,~b,~ \tau)$ and charginos respectively; (j=1,2 counts the
two different charginos). The relevant
$(Z,~ A^0)$-couplings appear in (\ref{Z-f-coupling}, \ref{H-f-coupling})
and (\ref{Z-chi1-coupling}, \ref{Z-chi2-coupling},
\ref{H-chi1-coupling}, \ref{H-chi2-coupling}).
 For the amplitudes  defined in (\ref{f-box-amplitude-A0}) we find
\bqa
A^{f-\rm box}_{+++}(A^0)&= & -
A^{f-\rm box}_{+++}(H)
~~, \label{A0-f-box+++} \\
A^{f-\rm box}_{++0}(A^0)&= & -
A^{f-\rm box}_{++0}(H)
~~, \label{A0-f-box++0}
\eqa
where (\ref{h0-f-box+++}, \ref{h0-f-box++0}) should be used
accompanied with the obvious  replacement
$m  \Rightarrow m_{A^0}$. For
the rest of the "basic" amplitudes in
(\ref{basic-A0-amplitudes}) we get
\bqa
&& A^{f-\rm box}_{+--}(A^0)=\frac{\sqrt{2} m_f}{\kappa \sqrt{Ys}}
\Bigg \{(t_Z+u_Z)(\kappa +u-t)Y D_{hZ}^f(t,u)
\nonumber \\
&& +(t_h+u_h)(\kappa +u-t) [2 s C_0^f(s)
-s u D_{hZ}^f(s,u)-st D_{hZ}^f(s,t) ]
\nonumber \\
&&- 2[t_h(t-\kappa)+u_Z m^2+Y][u_hC_h^f(u)+t_ZC_Z^f(t)]
\nonumber \\
&& +2[u_h(u+\kappa)+t_Z m^2+Y][u_ZC_Z^f(u)+t_hC_h^f(t)]
\Bigg \} ~~ ,\label{A0-f-box+--} \\
&& A^{f-\rm box}_{+-0}(A^0)= \frac{4 m_f}{\kappa \mz Y}
\Bigg \{ (t^2+u^2-2 m^2 \mzd)[s(t+u)C_0^f(s)-\kappa^2 C_{hZ}^f(s)]
\nonumber \\
&&- 2 m_f^2 \kappa^2 Y \tF^f(s,t,u)+2\mzd Y^2 D_{hZ}^f(t,u)
+[u^2(t+u)-m^2\mzd (3u-t)]E_1^f(s,u)
\nonumber \\
&& +[t^2(t+u)-m^2\mzd (3t-u)]E_1^f(s,t) \Bigg \}
~. \label{A0-f-box+-0}
\eqa

\vspace{1cm}
{\bf \underline{Mixed chargino box contribution.}}
This is determined by the diagrams in Fig.\ref{ggZA0-diag}c,d
which involve vector mixed Z-coupling to the two charginos, and those
of  Fig.\ref{ggZA0-diag}e,f containing axial mixed Z-couplings;
compare (\ref{Z-chi12-coupling}. Their complete contribution may
be written as
\bqa
F^{\tchi_1\tchi_2-\rm box}_{\lambda_1\lambda_2\lambda_Z}
(\gamma \gamma \to Z A^0 )&= &- \frac{\alpha e}{4 \pi}
\Big [ g_{v12}^Z g_{s12}^{A^0}
A^{\tchi_1 \tchi_2-\rm
box}_{\lambda_1\lambda_2\lambda_Z}(A^0,\mchil, \mchih )
\nonumber \\
& - & g_{a12}^Z g_{p12}^{A^0}
A^{\tchi_1 \tchi_2-\rm box}_{\lambda_1\lambda_2\lambda_Z}(
A^0,\mchil, -\mchih )
\Big ] ~, \label{chi12-box-amplitude-A0}
\eqa
where the two terms in the r.h.s. arise from the diagrams in
Fig.\ref{ggZA0-diag}(c,d) and (e,f) respectively;
and the Z- and $A^0$ couplings appear in
(\ref{Z-chi12-coupling}, \ref{H-chi12-coupling}). Thus
only the (c,d) diagrams need to be calculated.

Defining also
\bqa
Q_A &=& -4 s (\mchil^2+\mchih^2)+ 4 u_hu_Z+s^2+m^2s-\mzd s +4su
~, \nonumber \\
\bar Q_A&=& -\kappa^2 (m^2+\mzd)+(t+u)(t-u)^2+8\mzd Y ~,
\label{new-functions-3}
\eqa
\bqa
P_{tu}^+ &=& 2(t-u)Y\Big \{
-2s (\mchil^2+\mchih^2)(\mchil+\mchih)^2-(\mchil+\mchih)^2[2Y-s(t_Z+u_Z)]
\nonumber \\
&+&Y(t_Z+u_Z) \Big \} ~ , \nonumber \\
Q_{tu}^+ &=& 4(\mchil+\mchih)^2(t-u)\Big \{ 2 s^2
(\mchil^2-\mchih^2)^2+3s(\mchil^2+\mchih^2) Y+Y^2 \Big \} ~,
\nonumber \\
P_{tu}^- &=& 2 \Big \{ 2 s (t_Z+u_Z)(\mchil^2-\mchih^2)^2
-(\mchil-\mchih)^2 s (t^2+u^2-2 m^2\mzd )
\nonumber \\
&+& 2 (\mchil^2+\mchih^2) Y (t_Z+u_Z)-Y(t^2+u^2-2m^2\mzd)\Big \}
~ , \nonumber \\
Q_{tu}^- &=& 4\Big \{ s(t_Z+u_Z)(\mchil^2-\mchih^2)^2
+(\mchil^2+\mchih^2) Y (m^2-\mzd -2s)
\nonumber \\
&+& 2\mchil\mchih s Y-Y^2\Big \}
~, \label{new-functions-4}
\eqa
\bqa
 P_{su}^+ &=& 2 Y \Big \{
-2 (t-u)(\mchil+\mchih)^2 (\mchil^2+\mchih^2)
\nonumber \\
&+& (\mchil+\mchih)^2(2 u u_h +6 u_hu_Z+\kappa^2 -2\mzd t_h)
-u(t-u)(t_h+u_h) \Big \} ~ , \nonumber \\
 Q_{su}^+ & =& 4(\mchil+\mchih)^2 \Big \{
2 s (t-u)(\mchil^2-\mchih^2)^2 +3(t-u)Y(\mchil^2+\mchih^2)
\nonumber \\
&+& (Y+2u_Zu_h)[3(u^2-m^2 \mzd)-Y]\Big \} ~,
\nonumber \\
 P_{su}^- &=& -2Y \Big \{-2 (2s+t-u)(\mchil^2-\mchih^2)^2
-(t-u)(t_h+u_h)(\mchil-\mchih)^2
\nonumber \\
&+& 2 (\mchil^2+\mchih^2)(u^2-m^2\mzd -3Y)+
u[4 u t_h +t^2+3u^2-2\mzd(t+u)]\Big \} ~, \nonumber \\
 Q_{su}^-& =& -4 \Big \{ (\mchil^2-\mchih^2)^2
\Big [ (m^2+\mzd)[ 4 Y -6(u^2-m^2\mzd)]+6u^3-t u(t+u)
\nonumber \\
&+&m^2\mzd (u-5t)\Big ] +Y[Y-3(u^2-m^2\mzd )](\mchil^2+\mchih^2)
\nonumber \\
&-& 2 s u^2 (u^2-m^2\mzd)\Big \} ~. \label{new-functions-5}
\eqa
we find for the basic amplitudes (compare
(\ref{basic-A0-amplitudes}))
\bqa
&& A^{\tchi_1 \tchi_2-\rm box}_{+++}(A^0,\mchil, \mchih )=
\nonumber \\
&&  \frac{\sqrt{2}}{8\kappa \sqrt{s^3 Y}} \Bigg \{ \Bigg [
- 8s^2(t-u)\mchil [t_h+u_h-\kappa
 +4\mchil (\mchil+\mchih)]C_0^{111}(s)
\nonumber \\
&& +4 s (\mchil-\mchih)u_h \Big \{
(\kappa+u-t)[2(\mchil +\mchih)^2 +t_h]
-2Y \Big \} [C_h^{211}(u)+C_h^{122}(u)]
\nonumber \\
&&  -8 (\mchil+\mchih)Y
(\kappa-t_Z-u_Z)\Big \{ u_h [C_h^{211}(u)-C_h^{122}(u)]
 -u_Z [ C_Z^{211}(u)-C_Z^{122}(u)] \Big \}
\nonumber \\
&& -4s (\mchil-\mchih)u_Z \Big \{(\kappa+t-u)
[2(\mchil+\mchih)^2+u_h]
 -2Y\Big \}[ C_Z^{211}(u) + C_Z^{122}(u)]
\nonumber \\
&& -8s^2\mchil [4\mchil (\mchil+\mchih)+t_h+u_h-\kappa]
[Y+u^2-m^2\mzd
\nonumber \\
&&  -(t-u)(\mchil^2-\mchih^2)]D_{hZ}^{1211}(s,u)
 -s (\mchil-\mchih)(t-u)\Big \{ \kappa [s(\mchil+\mchih)^2+Y]
\nonumber \\
&& -(t_h+u_h)Y
 +(\mchil+\mchih)^2 Q_A \Big \}[D_{hZ}^{1221}(t,u)+D_{hZ}^{2112}(t,u)]
 +(\mchil+\mchih)\Big \{Y \bar Q_A
\nonumber \\
&&  -\kappa [s (\mchil-\mchih)^2+Y]
(Q_A-8\mchil\mchih s )+ s [\bar Q_A (\mchil^2+\mchih^2)
\nonumber \\
&& + 2 \mchil\mchih s \kappa^2  ] \Big \}
[D_{hZ}^{1221}(t,u)-D_{hZ}^{2112}(t,u)]-~ (t \leftrightarrow u)
\Bigg ] -(1 \leftrightarrow 2) \Bigg \} ~,
\label{A0-chi12-box+++}\\
&& A^{\tchi_1 \tchi_2-\rm box}_{+--}(A^0,\mchil, \mchih )=
\nonumber \\
&& \frac{1}{\kappa \sqrt{2 s Y^3}}\Bigg \{
8(\mchil+\mchih)s Y (\kappa+t-u)
[B_0(s,\mchil, \mchil)-B_0(s,\mchih, \mchih)]
\nonumber \\
&& -2(\mchil-\mchih)s \Big
\{(t-u-\kappa)Y[t_h+u_h+2(\mchil+\mchih)^2]
\nonumber \\
&& +2 (u-t-\kappa)(\mchil+\mchih)^2[Y+2s(t+u)]\Big \}
[C_0^{111}(s)+C_0^{222}(s)]
\nonumber \\
&& + 2(\mchil+\mchih)s \Big
\{(u-t+\kappa)Y[t_h+u_h+2(\mchil^2+\mchih^2)]+2(t-u+\kappa)
[Y(\mchil^2+\mchih^2)
\nonumber \\
&& +2s(\mchil^2-\mchih^2)^2 +s (t^2-m^2\mzd )+s u(t+u)]\Big \}
[C_0^{111}(s)-C_0^{222}(s)]
\nonumber \\
&&-8(\mchil-\mchih)(\mchil+\mchih)^2s\kappa \Big \{
\kappa (t-u)+t^2+u^2-2m^2\mzd \Big \}
[C_{hZ}^{121}(s)+C_{hZ}^{212}(s)]
\nonumber \\
&&-4(\mchil+\mchih)s (t+u)\kappa \Big \{
\kappa (t-u)+t^2+u^2-2m^2\mzd \Big \}[C_{hZ}^{121}(s)-C_{hZ}^{212}(s)]
\nonumber \\
&&- \frac{(\mchil-\mchih)}{2(u-t)}\Big \{ P_{tu}^+(\kappa+u-t)
+Q_{tu}^+(-\kappa+u-t) \Big \}
[D_{hZ}^{1221}(t,u)+D_{hZ}^{2112}(t,u)]
\nonumber \\
&&+ \frac{(\mchil+\mchih)}{2(u-t)}Y \Big \{ P_{tu}^-(\kappa+u-t)
+Q_{tu}^-(-\kappa+u-t) \Big \}
[D_{hZ}^{1221}(t,u)-D_{hZ}^{2112}(t,u)]
\nonumber \\
&&+ \Bigg [ 2(\mchil-\mchih) u_h \Big \{2 (\mchil+\mchih)^2
[-2m^2 (u^2-m^2\mzd)-4\mzd Y+3(t+u)Y
\nonumber \\
&&+ 2 u (uu_Z-tt_Z)]-Y(Y+m^2u_Z+tt_h)
\nonumber \\
&&+  \kappa [-2 (\mchil+\mchih)^2(Y+2u_Zu_h)+t_hY]\Big \}
[C_h^{211}(u)+C_h^{122}(u)]
\nonumber \\
&& +2(\mchil-\mchih)u_Z\Big \{
2(\mchil+\mchih)^2 [(2u_h+t)(u^2-m^2\mzd)-2\mzd u(u-t)
-u(t^2-m^2\mzd)]
\nonumber \\
&& +Y(Y+m^2t_Z+uu_h)+\kappa [-2
(\mchil+\mchih)^2(Y+2u_Zu_h)+u_hY]\Big \}[C_Z^{211}(u)+C_Z^{122}(u)]
\nonumber \\
&&-4(\mchil+\mchih)s u (Y+u^2 -m^2\mzd -\kappa u) \Big \{
u_h [C_h^{211}(u)-C_h^{122}(u)]
+u_Z[C_Z^{211}(u)-C_Z^{122}(u)]\Big \}
\nonumber \\
&&- \frac{(\mchil-\mchih)s}{2(u-t)}\Big \{ P_{su}^+(\kappa+u-t)
+Q_{su}^+(-\kappa+u-t)\Big \}
[D_{hZ}^{1211}(s,u)+D_{hZ}^{2122}(s,u)]
\nonumber \\
&& +\frac{(\mchil+\mchih)s}{2(u-t)}\Big \{ P_{su}^-(\kappa+u-t)
+Q_{su}^-(-\kappa+u-t)\Big \}
[D_{hZ}^{1211}(s,u)-D_{hZ}^{2122}(s,u)]
\nonumber \\
&& ~ -( t \leftrightarrow u ~~, ~~ \kappa \rightarrow -\kappa )
\Bigg ] \Bigg \} ~~ ,
\label{A0-chi12-box+--}\\
&& A^{\tchi_1 \tchi_2-\rm box}_{++0}(A^0,\mchil, \mchih )=
\nonumber \\
&& \frac{1}{2\kappa \mz s} \Bigg \{
8(\mchil-\mchih)s [s(t+u)-2(\mchil+\mchih)^2(t_Z+u_Z)]
[C_0^{111}(s)+C_0^{222}(s)]
\nonumber \\
&&+ 8(\mchil+\mchih)s [s(t+u)-2(\mchil^2+\mchih^2)(t_Z+u_Z)]
[C_0^{111}(s)-C_0^{222}(s)]
\nonumber \\
&& - 4(\mchil-\mchih) \Big \{ Y [(t+u)(m^2+\mzd)-4 m^2\mzd]
-2s (t_Z+u_Z)(\mchil^2+\mchih^2)(\mchil+\mchih)^2
\nonumber \\
&&+\frac{(\mchil+\mchih)^2}{2}(t_Z+u_Z)(4u_Zu_h+sm^2+ss_Z+4su)
+\frac{(\mchil-\mchih)^2}{2} s \kappa^2\Big \}\cdot
\nonumber \\
&& \cdot [D_{hZ}^{1221}(t,u)+D_{hZ}^{2112}(t,u)]
\nonumber \\
&&+ 8 (\mchil+\mchih)\mzd (t-u)[s (\mchil^2+\mchih^2)+Y]
[D_{hZ}^{1221}(t,u)-D_{hZ}^{2112}(t,u)]
\nonumber \\
&&+ \Bigg [ 4(\mchil-\mchih)[(t+u)(m^2+\mzd)-4m^2\mzd
-2(\mchil+\mchih)^2(t_Z+u_Z)]\cdot
\nonumber \\
&& \cdot \Big \{ u_h[C_h^{211}(u)+C_h^{122}(u)]
+u_Z[C_Z^{211}(u)+C_Z^{122}(u)] \Big \}
\nonumber \\
&&+ 8(\mchil+\mchih)\mzd (u-t) \Big \{
u_h[C_h^{211}(u)-C_h^{122}(u)] -u_Z[C_Z^{211}(u)-C_Z^{122}(u)] \Big \}
\nonumber \\
&& +4(\mchil -\mchih)s \Big \{
2(t_Z+u_Z)(\mchil^2+\mchih^2)(\mchil+\mchih)^2-2s m^2\mzd
\nonumber \\
&& +(\mchil^2+\mchih^2)[8m^2\mzd-(t+u)(m^2+3\mzd)]
+2\mchil\mchih [s^2+(\mzd-m^2)(s-2\mzd)]\Big \}\cdot
\nonumber \\
&& \cdot [D_{hZ}^{1211}(s,u)+D_{hZ}^{2122}(s,u)]
\nonumber \\
&& +4(\mchil +\mchih)s \Big \{
2(t_Z+u_Z)(\mchil^2-\mchih^2)^2 -2s m^2\mzd
 +(\mchil^2+\mchih^2)[ 8m^2\mzd
\nonumber \\
&& -(t+u)(m^2+3\mzd)]
-2\mchil\mchih [4m^2\mzd -(t+u)(m^2+\mzd)]\Big \}
 [D_{hZ}^{1211}(s,u)-D_{hZ}^{2122}(s,u)]
\nonumber \\
&&~~ + (u \leftrightarrow  t ) \Bigg ] \Bigg \} ~~,
\label{A0-chi12-box++0} \\
&& A^{\tchi_1 \tchi_2-\rm box}_{+-0}(A^0,\mchil, \mchih )=
\frac{1}{\kappa \mz Y} \Bigg \{
-(\mchil-\mchih)s \Big \{ 2 (\mchil+\mchih)^2
[\kappa^2-4\mzd (t_h+u_h)]
\nonumber \\
&& -(t+u)(t^2+u^2-2m^2\mzd)\Big \}[C_0^{111}(s)+C_0^{222}(s)]
+2(\mchil+\mchih)s \Big \{(\mchil-\mchih)^2 \kappa^2
\nonumber \\
&& -2 (t_Z+u_Z)(\mchil^2-\mchih^2)^2  +\mzd (2 m^2s_h+t^2+u^2)\Big \}
[C_0^{111}(s)-C_0^{222}(s)]
\nonumber \\
&& +(\mchil-\mchih)\Big \{ (t^2+u^2-2m^2\mzd)
[2(t_Z+u_Z)(\mchil+\mchih)^2-\kappa^2]\Big \}
[C_{hZ}^{121}(s)+C_{hZ}^{212}(s)]
\nonumber \\
&&- 2(\mchil+\mchih)\mzd (t_h+u_h)(t^2+u^2-2m^2\mzd)
[C_{hZ}^{121}(s)-C_{hZ}^{212}(s)]
\nonumber \\
&& +2 (\mchil-\mchih)u_h \Big \{ u (u^2-m^2\mzd)+u Y+m^2\mzd (t-u)
\nonumber \\
&& -2(\mchil+\mchih)^2(2 \mzd s_Z+u^2+t u) \Big \}
[C_h^{211}(u)+C_h^{122}(u)]
\nonumber \\
&& +2 (\mchil-\mchih)u_Z \Big \{ u (u^2-m^2\mzd)+u Y+m^2\mzd (t-u)
\nonumber \\
&& -2(\mchil+\mchih)^2(2 m^2 \mzd -4u \mzd +u^2+t u) \Big \}
[C_Z^{211}(u)+C_Z^{122}(u)]
\nonumber \\
&& +4(\mchil+\mchih)\mzd u (t_h+u_h) \Big \{
u_h[C_h^{211}(u)-C_h^{122}(u)]+u_Z[C_Z^{211}(u)-C_Z^{122}(u)]
\Big \}
\nonumber \\
&&-2(\mchil-\mchih)\Big \{ -2(t_Z+u_Z)(\mchil+\mchih)^2
[s(\mchil^2-\mchih^2)^2 +Y(\mchil^2+\mchih^2)]
\nonumber \\
&& +s\kappa^2 (\mchil^2-\mchih^2)^2+ 2(\mchil+\mchih)^2
[u_h\mz^6+\mz^4(m^4-u^2+2m^2s-s u)
\nonumber \\
&&+u u_h\mzd (3s+m^2)+s u^2 s_h]+s u (u^3+tu^2-3\mzd m^2 u
+m^2\mzd t)\nonumber \\
&& +\kappa^2 Y (\mchil^2+\mchih^2)\Big \}
 [D_{hZ}^{1211}(s,u)+D_{hZ}^{2122}(s,u)]
\nonumber \\
&& +2(\mchil+\mchih)\Big \{-2 (\mchil^2-\mchih^2)^2
[m^2 (u^2-\mz^4)-m^4u_Z-2m^2\mzd s-\mzd (s-u)^2
\nonumber \\
&&+s u (s-u)+\mz^4(s+u)]-(\mchil-\mchih)^2
\{2 \mzd m^2 (2\mzd m^2-t^2+3u^2)
\nonumber \\
&& +(m^2+\mzd)[m^2\mzd(t-7u)+2u^3+3tu^2+t^2u]
-2 u^2(t+u)^2\}
\nonumber \\
&& -2\mzd Y (\mchil^2+\mchih^2)(t_h+u_h)
-2 s u \mzd (u^2-m^2\mzd -2m^2 u_Z)\Big \}
[D_{hZ}^{1211}(s,u)-D_{hZ}^{2122}(s,u)]
\nonumber \\
&&+ 2(\mchil+\mchih)\mzd (t-u)
[s(\mchil^2-\mchih^2)^2+ (\mchil^2+\mchih^2) Y]
[D_{hZ}^{1221}(t,u)-D_{hZ}^{2112}(t,u)]
\nonumber \\
&&-(\mchil-\mchih)\Big \{[(\mchil^2-\mchih^2)^2 s
+(\mchil^2+\mchih^2) Y][\kappa^2-2(t_Z+u_Z)(\mchil+\mchih)^2]
\nonumber \\
&& - 2\mzd Y [s (\mchil+\mchih)^2+Y]\Big \}
[D_{hZ}^{1221}(t,u)+D_{hZ}^{2112}(t,u)]
~~+(t \leftrightarrow u) \Bigg \} ~.
\label{A0-chi12-box+-0}
\eqa

\clearpage
\newpage

\clearpage
\newpage

\clearpage
\newpage

\clearpage
\newpage

\clearpage
\newpage

\end{document}